\documentclass[runningheads]{llncs}

\usepackage[lcgreekalpha]{stix}
\usepackage[numbers]{natbib}
\usepackage{amsmath}
\usepackage{siunitx}
\usepackage{subfig}
\usepackage{mathtools}
\usepackage{lineno}
\usepackage{multirow}
\usepackage{booktabs}
\usepackage{algorithm}
\usepackage{transparent}
\usepackage{tabulary}
\usepackage{makecell} \usepackage{algpseudocode}
\usepackage{graphicx}
\usepackage{pgfplots}
\usepackage{xcolor}
\usepackage{longtable}
\usepackage[hidelinks,colorlinks=true,citecolor=blue,linkcolor=red,urlcolor=green,pdfstartview=]{hyperref}
\usepackage[locale=DE]{siunitx}
\usepackage{nicefrac}
\usepackage{adjustbox}
\usepackage{pdflscape}
\usepackage[nolist,nohyperlinks]{acronym}

\DeclareSIUnit\M{M}

\usepackage{dsfont}

\usepackage{bm}

\usepackage{subfig}
\usepackage{transparent}
\graphicspath{{figures/}}
\usepackage{caption}

\usepackage{pgfplots}

\usepgfplotslibrary{external}

\usepackage{soul}

\pgfplotsset{compat=1.18}

\begin{acronym}[MPTPS]
  \acro{CBF}{cerebral blood flow}
  \acro{NVU}{neurovascular unit}
  \acro{ECs}{endothelial cells}
  \acro{NGVU}{neuro-glial-vascular unit}
  \acro{Glu}{glutamate}
  \acro{ATP}{adenosine triphosphate}
  \acro{NO}{nitric oxide}
  \acro{MLCK}{myosin light-chain kinase}
  \acro{MLCP}{myosin light-chain phosphatase}
  \acro{CaM}{calmodulin}
  \acro{EDHF}{endothelium-derived hyperpolarizing factor}
  \acro{ET-1}{endothelin-1}
  \acro{TRP}{transient receptor potential}
  \acro{NMDA}{N-methyl-D-aspartate}
  \acro{AMPA}{$\alpha$-amino-3-hydroxy-5-methyl-4-isoxazolepropionic acid}
  \acro{GABA}{$\gamma$-aminobutyric acid}
  \acro{IP$_3$}{Inositol trisphosphate}
  \acro{CICR}{Calcium-induced calcium release}
  \acro{VOCC}{voltage-operated calcium channels}
  \acro{KIR}{inward-rectifying potassium channels}
  \acro{NBC}{sodium bicarbonate pump}
  \acro{KCC1}{potassium chloride co-transporter pump}
  \acro{cGMP}{cyclic guanosine monophosphate}
  \acro{cAMP}{cyclic adenosine monophosphate}
  \acro{O$_2$}{oxygen}
  \acro{CO$_2$}{carbon dioxide}
  \acro{CPP}{cerebral perfusion pressure}
  \acro{MAP}{mean arterial pressure}
  \acro{ICP}{intracranial pressure}
  \acro{AP}{arterial pressure}
  \acro{TCA}{tricarboxylic acid}
  \acro{ROS}{reactive oxygen species}
  \acro{NADH}{nicotinamide adenine dinucleotide (hydrogen)}
  \acro{PUFAs}{polyunsaturated fatty acids}
  \acro{SEM}{scanning electron microscopy}
  \acro{DVC}{dorsal vagal complex}
  \acro{NTS}{nucleus tractus solitarius}
  \acro{PMCa}{plasma membrane Ca$^{2+}$ ATPase}
  \acro{NOS}{nitric oxide synthases}
  \acro{nNOS}{neuronal nitric oxide synthases}
  \acro{eNOS}{endothelial nitric oxide synthases}
  \acro{L-Arg}{L-Arginine}
  \acro{NADPH}{nicotinamide adenine dinucleotide phosphate (hydrogen)}
  \acro{NADP}{nicotinamide adenine dinucleotide phosphate}
  \acro{SMCs}{smooth muscle cells}
  \acro{EPSC}{excitatory post-synaptic current}
  \acro{IPSC}{inhibitory post-synaptic current}
  \acro{ER}{endoplasmic reticulum}
  \acro{PSP}{post-synaptic potential}
  \acro{PICA}{posterior inferior cerebellar artery}
\end{acronym}

\begin{document}

\title{A 3D-1D-0D Multiscale Model of the Neuro-Glial-Vascular Unit for Synaptic and Vascular Dynamics in the Dorsal Vagal Complex}

\titlerunning{Synaptic and Vascular Dynamics in the NGVU of the DVC}

\author{ Alexander Hermann$^{^*}$\inst{1,2}\orcidID{0000-0002-9731-3286}    \and
 Tobias Köppl\inst{3}\orcidID{0000-0003-3548-2807}         \and
 Andreas Wagner\inst{4}\orcidID{0000-0002-1622-846X}       \and
 Arman Shojaei\inst{1}\orcidID{0000-0001-8638-8285}        \and \\
 Barbara Wohlmuth\inst{4}\orcidID{0000-0001-6908-6015}     \and
 Roland Aydin\inst{1}\orcidID{0000-0002-9542-9146}         \and \\
 Christian J. Cyron\inst{1,5}\orcidID{0000-0002-0715-6484}     \and 
 Roustem Miftahof\inst{5}\orcidID{0000-0001-8264-0885}
}
\authorrunning{A. Hermann, T. Köppl, A. Wagner et al.}

\institute{    Institute of Material Systems Modeling, Helmholtz-Zentrum Hereon, Max-Planck-Str. 1, 21502 Geesthacht, Germany
    \and
    CAU Innovation GmbH, Fraunhoferstraße 13, 24118 Kiel, Germany \and 
    Fraunhofer-Institut FOKUS, Kaiserin-Augusta-Allee 31, 10589 Berlin, Germany \and
    Department of Mathematics, Technical University of Munich, Garching, Germany \and
    Institute for Continuum and Material Mechanics, Hamburg University of Technology, Eissendorfer Strasse 42, 21073 Hamburg, Germany
    \\
    $^*$ Corresponding author: \email{alexander.hermann@hereon.de}
}
\maketitle
\begin{abstract}
Cerebral blood flow regulation is critical for brain function, and its disruption is implicated in various neurological disorders. Many existing models do not fully capture the complex, multiscale interactions among neuronal activity, astrocytic signaling, and vascular dynamics—especially in key brainstem regions. In this work, we present a 3D-1D-0D multiscale computational framework for modeling the neuro-glial-vascular unit (NGVU) in the dorsal vagal complex (DVC). Our approach integrates a quadripartite synapse model—which represents the interplay among excitatory and inhibitory neurons, astrocytes, and vascular smooth muscle cells—with a hierarchical description of vascular dynamics that couples a three-dimensional microcirculatory network with a one-dimensional macrocirculatory representation and a zero-dimensional synaptic component. By linking neuronal spiking, astrocytic calcium and gliotransmitter signaling, and vascular tone regulation, our model reproduces key features of functional hyperemia and elucidates the feedback loops that help maintain cerebral blood flow. Simulation results demonstrate that neurotransmitter release triggers astrocytic responses that modulate vessel radius to optimize oxygen and nutrient delivery. This integrated framework, to our knowledge the first model to combine these elements for the NGVU in the DVC, provides a robust and modular platform for future investigations into the pathophysiology of cerebral blood flow regulation and its role in autonomic control, including the regulation of stomach function.
\end{abstract}

\section{Introduction}\label{sec:introduction}
\begin{figure}[htb]
    \centering
    \includegraphics[width=\linewidth]{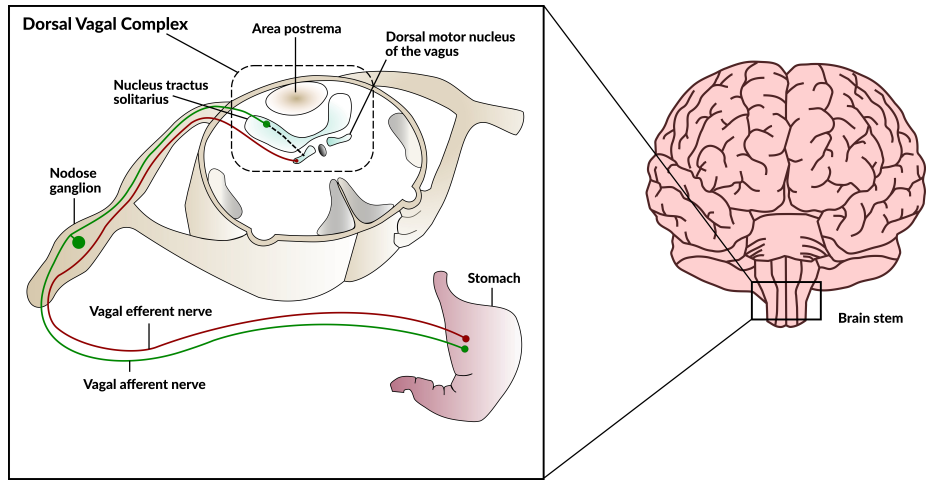}
    \caption{Schematic representation of the dorsal vagal complex. A detailed view of the dorsal vagal complex is depicted, including the stomach, vagal afferent and efferent nerves, nodose ganglion, nucleus tractus solitarius, area postrema, and the dorsal motor nucleus of the vagus (adapted from~\citep{waise2018metabolic}).}
    \label{fig:anatomy}
\end{figure}

\begin{figure}[htbp!]
    \centering
                                        \includegraphics[width=\linewidth]{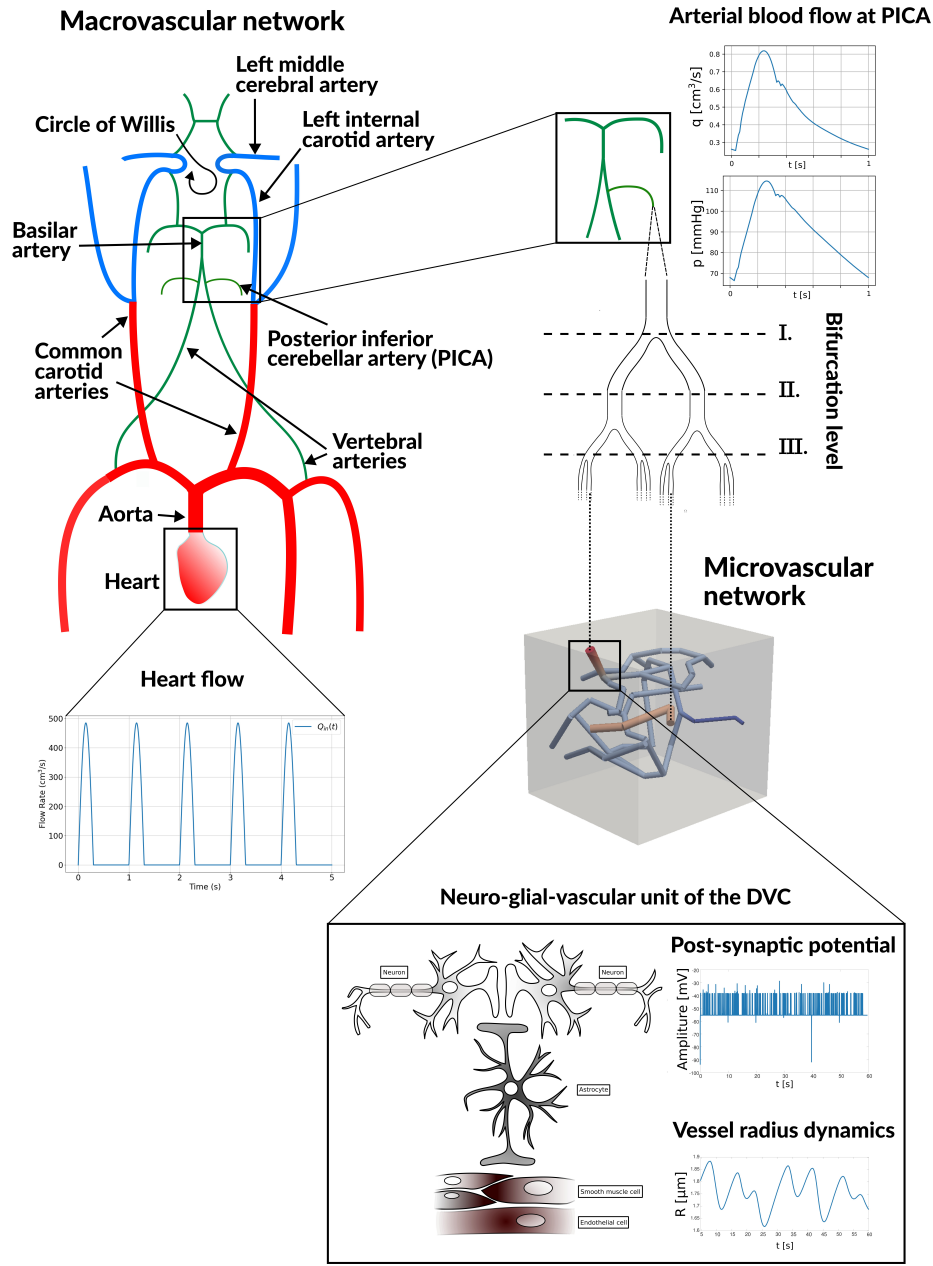}
                \caption{Graphical Abstract. Schematic overview of our novel multiscale modeling framework for the \ac{NGVU} in the \ac{DVC}. The top panel shows the heart generating a pulsatile blood flow at one beat per second that is transmitted through major arteries including the common carotid, vertebral, and internal carotid arteries and collected in the circle of Willis. The arterial blood then enters the \ac{PICA} to supply the \ac{DVC} via a series of symmetric bifurcations in the mesoscale arterial tree. The bottom panel shows a detailed view of the local neuro-glial-vascular unit where the quadripartite synapse model captures interactions among excitatory and inhibitory neurons, astrocytes, and vascular smooth muscle cells. This integration dynamically regulates vessel radius and optimizes oxygen delivery to surrounding tissue through a one-dimensional to three-dimensional coupling approach.}
    \label{fig:schematic}
\end{figure}

Understanding perception, cognition, learning, memory and consciousness in the brain are the prime objectives of neuroscience. Until now, research on the brain has been dominated by both experimental and theoretical reductionism, which emphasizes detailed knowledge of its structure and function at a high-resolution level. Successful examples of this approach include a computationally intensive reconstruction of morphological constituents in a cubic millimeter of the human temporal cortex at a nanometer scale~\citep{shapson2024petavoxel}; the uncovering of cellular diversity at the transcriptomics level~\citep{schaeffer2021revisiting}; real-time monitoring of the dynamics of synaptic plasticity using genetically encoded, intensiometric fluorescence indicators~\citep{son2024real}. Although purely reductive strategies provide amazing insights into the functional intricacies of biological matter, they unequivocally neglect the context in which these processes operate. This severely limits the ability of reductionism to explain how and why the brain behaves as it does. Notable examples of the conceptual and expensive shortcomings are The Blue Brain, The European Human Brain, and The Cajal Blue Brain Projects—research initiatives that intended but failed to create a digital replica of functional mouse and human brains. The reductive trend continues to prevail in \emph{in vitro} and \emph{in vivo} studies of brain function under physiological and disease states. It is attractive from experimental and theoretical perspectives to investigate a single target in isolation, such as dopamine neurotransmission in the pathogenesis of Parkinson’s disease, neuronal activity in the pathophysiology of epilepsy, or \ac{CBF} in stroke and vascular dementia patients. However, being deterministic in nature, the underlying physiological processes run at multi-hierarchical levels and cannot be accounted for by reductionism. This shift has led to a change in the paradigm of studying the finely parcellated brain, moving from a reductionist approach to an integrative analysis at different levels of its organization. 
\\
\indent The concept of the \ac{NVU} as a functional element of the brain was proposed and formalized in 2001 at the inaugural Stroke Progress Review Group meeting of the National Institute of Neurological Disorders and Stroke~\citep{iadecola2017neurovascular}. Initially, it was defined morphologically as a hetero-cellular structure composed of neurons, \ac{ECs}, vascular \ac{SMCs}, and the extracellular matrix. The \ac{NVU} was assigned \emph{a priori} properties of functional homogeneity and generality throughout the brain~\citep{schaeffer2021revisiting}. Over the years, the concept has evolved from its “canonical” form to a \ac{NGVU}, incorporating additional cellular constituents such as astrocytes and pericytes, to highlight the multi-functional, -dimensional, and -cellular coordinated function of the brain. Recent single-cell RNA-seq analyses have proven the molecular heterogeneity of brain cells~\citep{tripathy2017transcriptomic}. This has inferred the functional diversity of \ac{NGVU}s that depends on topological cell arrangements, regional specialization, and the metabolic demands of different brain areas. The most investigated function of the \ac{NVU} is neurovascular coupling, i.e., the matching of blood flow to neuronal metabolic demands, also known as functional hyperemia. Although the large vessel architecture of the brain is established at birth, the intracerebral vascular network grows postnatally, known as capillary angiogenesis, to match cortical expansion~\citep{marin2012human}. Two main cell types, \ac{ECs} and pericytes, embedded in the basal lamina, constitute the wall of capillaries. \ac{ECs} linked via gap junctions create a smooth luminal syncytium of the capillaries. These are ensheathed externally by pericytes. These cells have variable morphology ranging from circumferential, longitudinal to amoeboid depending on the location along the vascular tree~\citep{hartmann2015pericyte}. Pericytes mediate precise local changes of capillary diameter in response to excitatory/inhibitory signals arriving from the neurons and astrocytes. Coupled via gap junctions, pericytes and \ac{ECs} together sustain the conduction of locally induced signals along the capillary and arteriolar networks and thus regulate \ac{CBF} over space and time~\citep{hall2014capillary,longden2017capillary,cai2018stimulation,rungta2018vascular,emerson2000electrical,emerson2000endothelial,otani2023computational, ii2020multiscale}.
\\
\indent During capillary angiogenesis, astrocytes undergo massive proliferation. The basal lamina of blood vessels and the astrocyte-secreted basal lamina merge to form a common cerebrovascular basal lamina. This sets close contacts between the \ac{ECs}, pericytes, and astrocytes. Mature astrocytes vary in density by brain region and display regional heterogeneity in morphology~\citep{emsley2006astroglial,bayraktar2020astrocyte}. Interconnected via gap junctions, they form a topologically distinct network that ensheathes through astrocytic perivascular endfeet processes, 99.7\% of the capillary network and, concurrently, through its peri-synaptic processes, enwraps pre- and post-synaptic terminals of neurons, forming \emph{tripartite synapses}~\citep{bushong2002protoplasmic,kofuji2004potassium,tsai2012regional,mathiisen2010perivascular}. These provide efficient connections between the neuronal, astrocytic, and vascular networks and drive neurovascular coupling within the \ac{NGVU}~\citep{mishra2017binaural,hautefort2019endothelial,presa2020vasculo}. Astrocytes, \ac{ECs}, and pericytes sense signaling within \ac{NGVU}s and adjust \ac{CBF} according to metabolic demands. The release of \ac{Glu} and \ac{ATP} at the tripartite synapse triggers an influx of extracellular Ca$^{2+}$ to astrocytes. The increase in cytosolic Ca$^{2+}$ activates the release of gliatransmitters. Their biological effects depend on the site of action, the types of activated receptors, and the ligand concentration. Thus, \ac{NO} and epoxyeicosatrienoic acid have been shown to produce strong capillary vasodilation. 20-hydroxyeicosatetraenoic acid, prostaglandin E2, and Ca$^{2+}$ ions have potent vasoconstrictive~\citep{biesecker2016glial,magaki2018glial,thakore2021brain}, while astrocyte-derived \ac{ATP} and its metabolite adenosine exhibit bidirectional effects on the capillary wall~\citep{zaritsky2000targeted,filosa2006local,longden2015vascular}. Intriguingly, astrocytes can also perceive changes in capillary blood pressure and, in turn, amend neuronal firing activity~\citep{kim2015gintonin,kim2016cortical}. This vasculo-neuronal hypothesis may present a mechanism through which body state can dictate, or at least modulate, brain function~\citep{moore2008hemo,filosa2016beyond,presa2020vasculo}. \ac{ECs} react to changes in intracapillary pressure/flow and wall shear stress by releasing \ac{NO}, endothelial factor-1, prostacyclin, and proteases. These cause pericyte relaxation, break up the extracellular matrix, induce vascular remodeling, and neovascularization that alter \ac{CBF} to meet urgent requirements.
\\
\indent The study of nonlinearity, sensitivity to initial and state variables, stochasticity, and chaotic behavior of the \ac{NGVU} is central to understanding and identifying the key determinants of the (patho)physiology of the brain. Mathematical modelling of \ac{NGVU}s has witnessed impressive advances within the last decade due to an explosion in the amount and quality of experimentally available data. Two broad classes of models have emerged, i.e., i) neuron abstract, and ii) neuron morphologically and physiologically extended network models. While the first class captures essential architectural and electrical neuronal features, the second class accentuates the internal physiological signal processing underlining the repertoire of electrical activity observed \emph{in vivo}. The number of publications on the subject is continuously increasing and it is beyond our scope to cite and evaluate critically each model and their produced results. We rather consider it the reader’s responsibility to stay updated on progress in the field while exercising thorough impartial scientific judgment regarding the veracity, biological plausibility, and significance of the models. From our literature review, we conclude that at the time of writing there are no models of the \ac{NGVU} that capture its role as a fundamental morpho-functional unit of the human brain. Current models, without exception, are based on fragmentary analyses of constitutive components, and not a holistic representation of the unit with its complex biology. Consequently, these models do not support investigations or provide answers to questions about the brain’s perception and cognition in health and disease~\citep{el2021modelling, chen2024modelling}. To address this gap, we introduce a 3D-1D-0D multiscale computational tool designed to simulate the dynamics of the \ac{NGVU} in a more integrated manner. Our model is built upon established biological principles and existing research, with validation currently focused on qualitative alignment with these foundations. However, due to the complexity of the system and the limitations of available experimental data, full quantitative validation remains a future direction to enhance the model’s predictive accuracy.
\\
\indent Within this study, we direct our attention to the \ac{DVC}, a key brainstem structure located in the medulla oblongata. The \ac{DVC} plays a pivotal role in autonomic control by integrating visceral sensory and motor signals that regulate parasympathetic outflow to the cardiovascular, respiratory, and gastrointestinal systems~\citep{mahdi2013modeling}. It works closely with neighboring nuclei, including the \ac{NTS}, to coordinate reflexes such as baroreception (blood pressure regulation) and gut motility. Because the \ac{DVC} must respond rapidly to changing homeostatic demands, studying its local neuro-glial-vascular interactions is especially relevant~\citep{depitta2011tale}. Fluctuations in autonomic activity require fine-tuned mechanisms to ensure adequate blood flow and metabolic support, making the \ac{DVC} an ideal site for investigating the core principles of multicellular coordination in the brainstem~\citep{coggan2018norepinephrine, choi2019synchronization}. To facilitate a clearer understanding of our multi-scale modeling framework, Figure~\ref{fig:anatomy} provides a schematic illustration that depicts the stepwise construction of the model—from the macrovascular network (including the carotid, vertebral arteries and the circle of Willis), through the mesoscale arterial tree, to the microvascular unit representing a capillary vessel in the \ac{DVC}, and finally the integration of the quadripartite synapse model with dynamic vessel radius regulation. By focusing on this region, we also hope to shed light on how disruptions in these processes could contribute to disorders affecting autonomic functions.
\\
\indent The aim of this paper is to provide a modular framework for constructing a comprehensive \emph{in silico} model of the \ac{NGVU}, which represents the first integrated model of its kind for the \ac{DVC}. We begin with a gliovascular module in which astrocytic glia supports the function of associated cerebrovascular segments. A brief background of the major biological components that justify the working assumptions underlying our mathematical modelling of the essential physiological processes is presented in Section~\ref{sec:biological_foundations}. This is followed by a description of a spatiotemporal computational model of the \ac{NGVU} in Section~\ref{sec:NGVU_Model}, which includes discussions of the blood flow model, the quadripartite synapse model, the components of the neurovascular coupling model, and their integration. Numerical simulation results are presented in Section~\ref{sec:results}. Section~\ref{sec:discussion} discusses the findings and Section~\ref{sec:conclusion} summarizes the contributions and suggests directions for future research. For a concise visual summary of our approach and its novelty, see the Graphical Abstract in Figure~\ref{fig:schematic}.
\newpage
\section{Biological Foundations of Neuro-Glial-Vascular Interactions}
\label{sec:biological_foundations}
\subsection{Anatomical and Morphological Components of the NGVU}
The \ac{NGVU} consists of structurally and functionally interconnected components essential for modeling brain physiology. Key elements include neurons, glial cells (especially astrocytes), \ac{SMCs}, \ac{ECs}, and blood vessels. These components work in concert to regulate cerebral blood flow, vascular tone, and neural activity, making their anatomical, morphological, and physiological characteristics vital for constructing a biologically accurate mathematical model.
\\
\indent Neurons, the fundamental morphological and functional units of the brain, number approximately $10^{11}$ in the adult human brain. Structurally, they comprise the soma, axon(s), dendrites, nerve terminals, and spine buds~\citep{heinrich2014loffler}. The soma, measuring $\SI{3}{\micro\metre}$ to $\SI{18}{\micro\metre}$ in diameter~\citep{chudler2013brain}, contains essential organelles such as the nucleus, \ac{ER}, mitochondria, and ribosomes, which sustain neuronal functionality~\citep{zilles2011anatomie,ariyo2024anatomy}. Axons, cylindrical extensions of the soma, have diameters ranging from $\SI{0.1}{\micro\metre}$ to over $\SI{10}{\micro\metre}$ and lengths of $\SI{20}{\micro\metre}$ to $\SI{200}{\micro\metre}$, with collateral branches exceeding $\SI{10}{\micro\metre}$~\citep{lee2019along,samuels2003office}. They end in nerve terminals or terminal boutons, about $1$--$\SI{2}{\micro\metre}$ in diameter~\citep{zilles2011anatomie}. Axons can be myelinated or unmyelinated, depending on the presence of a myelin sheath \citep{zilles2011anatomie,samuels2003office}. Dendrites (from Greek $\delta \acute{\epsilon} \nu \delta \rho o \nu$, meaning \emph{tree}) are shorter, thinner processes branching from the soma, characterized by spine buds (0.5--$\SI{2}{\micro\metre}$ in diameter) connected by a spine neck ($0.04$--$\SI{1}{\micro\metre}$ long)~\citep{harris1994dendritic}. Unlike axons, dendrites have a poorly developed cytoskeleton and fewer neurofilaments but are rich in microtubules and microfilaments. The nerve terminals and spine buds are key sites for synaptic connections (from Greek $\sigma \acute{\upsilon} \nu \alpha \psi \iota \varsigma$, meaning \emph{conjunction})~\citep{reichardt1983molecular}, consisting of a presynaptic terminal, a synaptic cleft ($20$--$\SI{30}{\nano\metre}$), and a postsynaptic terminal. These connections form between neurons, astrocytes, or combinations thereof via dendritic, axonal, axo-dendritic, and axo- or dendro-somatic synapses~\citep{zilles2011anatomie}.
\\
\indent Astrocytes, oligodendrocytes, microglia, and radial cells belong to the class of glial cells~\citep{heinrich2014loffler, zilles2011anatomie}. Astrocytes are star-shaped cells with a relatively small soma measuring $10$--$\SI{20}{\micro\metre}$~\citep{zilles2011anatomie}. Six or more major branches extend from the soma, further dividing into branchlets and leaflets, with cross-sectional diameters ranging from $10$ to $\SI{100}{\nano\metre}$~\citep{chai2017neural,baldwin2024astrocyte}. Morphologically, astrocytes are categorized into two types: fibrous, found in white matter, and protoplasmic, located in gray matter. Both types feature fewer organelles and unbranched processes up to $\SI{300}{\micro\metre}$ in length~\citep{ariyo2024anatomy,goenaga2023calcium}. The process endings, known as endfeet, form astro-dendritic, astro-axonal, and neuron-smooth muscle synapses within the NGVU~\citep{kacem1998structural,garman2011histology,simard2003signaling}. Astrocytes possess a well-developed \ac{ER}, which serves as the primary intracellular Ca$^{2+}$ store, along with numerous mitochondria and an extensive microfilament network~\citep{shigetomi2016probing,sherwood2021astrocytic}.
\\
\indent Smooth muscle cells form the walls of cerebral blood vessels~\citep{zilles2011anatomie}. These fusiform-shaped cells measure approximately $\SI{20}{\micro\metre}$ in length and $\SI{6}{\micro\metre}$ in width, playing a crucial role in regulating vascular tone and providing mechanical elasticity to the vascular wall~\citep{hill2014vascular,wight2008arterial}. \ac{SMCs} form a syncytium through multiple gap junctions and basal laminae invaginations~\citep{zilles2011anatomie}. Their contractile apparatus comprises contractile proteins, regulatory proteins, and a cytoskeleton. The primary contractile proteins are actin and myosin, organized into thin and thick filaments (myofilaments). Thin filaments, approximately $\SI{7}{\nano\metre}$ in diameter, consist mainly of actin, while thick filaments, measuring $12$–$\SI{15}{\nano\metre}$ in diameter, are predominantly composed of myosin. Regulatory proteins such as \ac{MLCK}, \ac{MLCP}, \ac{CaM}, and caldesmon, along with the intracellular arrangement of myofilaments, constitute the functional contractile apparatus~\citep{zilles2011anatomie,chen2020biomechanical}.
\\
\indent The luminal surface of blood vessels is lined with a monolayer of \ac{ECs}, forming a selective semi-permeable membrane that regulates the movement of small molecules (O$_2$, CO$_2$, H$_2$O) and ions (Ca$^{2+}$, K$^+$) between the blood and brain parenchyma. These cells also modulate cerebral blood flow, vascular tone, inflammation, thrombosis, and adhesion. Endothelial dysfunction is a key factor in the pathogenesis of cerebral small vessel disease~\citep{zilles2011anatomie,sweeney2018blood}. \ac{ECs} synthesize and release vasoactive mediators, including \ac{NO}, \ac{EDHF}, eicosanoids, and \ac{ET-1}, to regulate vascular tone. \ac{NO} and \ac{EDHF} act as potent vasodilators by influencing the contractile apparatus and membrane hyperpolarization of \ac{SMCs}, while eicosanoids and \ac{ET-1} function as vasoconstrictors, inducing intense smooth muscle contractions~\citep{furchgott1980obligatory,longden2017capillary}.
\\
\indent Calcium ions act as crucial secondary messengers in regulating endothelial cell processes. Under physiological conditions, \ac{ECs} maintain a low cytosolic Ca$^{2+}$ concentration ($\SI{100}{\nano\mol\per\litre}$) compared to the extracellular concentration of $1.5$--$\SI{2}{\milli\mol\per\litre}$ in the brain. Cytoplasmic Ca$^{2+}$ is essential for the synthesis and release of endothelium-derived factors, including \ac{NO}. Agonist-driven \ac{NO} release, such as by acetylcholine, is mediated by intracellular Ca$^{2+}$ oscillations. Mechanical stimuli sensed by mechanosensitive ion channels elevate intracellular Ca$^{2+}$ levels, leading to endothelium-dependent smooth muscle hyperpolarization and relaxation. Additionally, reactive oxygen species produced by \ac{ECs} can activate \ac{TRP} channels, triggering Ca$^{2+}$ influx and dilating cerebral arteries~\citep{tran2000calcium,ando2013flow}.

\subsection{Transmitter, Ion Channels and Electrical Signal Transduction}
The \ac{NGVU} relies on various electrochemical processes to facilitate neural communication and vascular regulation. These processes are driven by the coordinated action of neurotransmitters, ion channels, and second messengers, which underlie the generation and propagation of signals across neurons, astrocytes, and vascular components. Understanding the dynamics of these signaling mechanisms is essential for modeling the complex interactions within the \ac{NGVU}, as they regulate neural activity, synaptic transmission, and vascular tone in both physiological and pathological conditions.
\\
\indent Transmitters are essential for the electrochemical processing of information at chemical synapses in the \ac{NGVU}. \ac{Glu}, the primary excitatory neurotransmitter, is used by nearly 90\% of neurons and accounts for 80--90\% of brain synapses. Synthesized \emph{de novo} from glucose or via glutamine recycling, \ac{Glu} is stored in vesicles and released by exocytosis~\citep{heinrich2014loffler}. In the synaptic cleft, \ac{Glu} binds to ionotropic receptors (\ac{NMDA}, \ac{AMPA}, and kainate) to alter ion channel permeability or to metabotropic receptors, which regulate intracellular signaling via G-proteins~\citep{sahlender2014we, montana2006vesicular}. \ac{GABA}, the primary inhibitory neurotransmitter, is synthesized via the \ac{GABA}-shunt and operates through ionotropic \ac{GABA}a and metabotropic \ac{GABA}b receptors. \ac{GABA}a activation causes Cl$^-$ influx and hyperpolarization, while \ac{GABA}b reduces Ca$^{2+}$ influx and increases K$^+$ efflux, inhibiting cell excitability~\citep{heinrich2014loffler, kubota1994three}. \ac{NO}, a gaseous transmitter synthesized on demand by \ac{NO} synthases, diffuses freely due to its solubility in fat and water~\citep{alderton2001nitric, forstermann1998expressional}. Despite its short half-life ($\SI{2}{\milli\second}$ intravascularly and up to $\SI{2}{\second}$ extravascularly), \ac{NO} mediates significant physiological effects, including smooth muscle relaxation and vasodilation~\citep{dormanns2016role, garthwaite1995nitric}. Endothelial cell-derived \ac{NO} primarily drives vasorelaxation during rest, while neuronal \ac{NO} dominates during activation, influencing arteriolar diameter by diffusing into \ac{SMCs}~\citep{dormanns2016role}. \ac{NO} release, triggered by wall shear stress, delays the return of arteriolar radius to baseline~\citep{thomas2001biological}. Together, \ac{Glu} and \ac{NO} mediate excitatory synaptic signaling and vascular regulation, while the interplay with inhibitory \ac{GABA} ensures neural and vascular homeostasis in the \ac{NGVU}~\citep{bezzi2001neuron, heinrich2014loffler, dormanns2016role}.
\\
\indent In addition to signaling molecules, second messengers play a crucial role in intracellular signal transmission by forming after the binding of signaling molecules to membrane-bound or cytosolic receptors~\citep{heinrich2014loffler}. \ac{IP$_3$}, a cyclic compound derived from membrane phospholipids during the \ac{IP$_3$} pathway, triggers Ca$^{2+}$ release from the \ac{ER}~\citep{heinrich2014loffler}. \ac{ATP}, apart from serving as an energy source, acts as a second messenger by binding to purinergic receptors and is co-localized with other neurotransmitters~\citep{heinrich2014loffler}. In astrocytes, \ac{ATP} facilitates Ca$^{2+}$ wave propagation and is released in a Ca$^{2+}$-dependent manner or through lysosome exocytosis, hemichannels, and anion channels~\citep{xiong2018stretch,lee2015ca2+,kang2008connexin}. Cellular membranes, composed of amphiphilic lipid bilayers with hydrophilic head groups facing outward and hydrophobic tails inward, serve as structural barriers enabling the exchange of substances between cells and their environment~\citep{heinrich2014loffler, brandes2019physiologie}. Ion channels, integral membrane proteins, facilitate the passage of ions like Na$^+$, K$^+$, Ca$^{2+}$, and Cl$^-$ across the lipid bilayer through selective water-filled pathways, achieving transport rates of 10$^7$--10$^8$ ions per second and forming the basis of cellular excitation~\citep{brandes2019physiologie}.
\\
\indent The movement of ions is driven by two primary forces: the concentration gradient and the potential difference. Passive transport occurs along an electrochemical gradient, with the ion current determined by the channel's conductivity and open probability. Greater membrane potential amplitude increases the driving force and current amplitude. Depending on channel type, ion concentration, and temperature, ion channel conductivities range from $1$ to $10^{-10}$~S, while specific membrane resistances vary from $10^{12}$ to $10^9$~$\Omega$~\citep{brandes2019physiologie}. Neurons of the NGVU model express voltage-dependent Na$^+$ channels, L-type and N-type Ca$^{2+}$ channels, Cl$^{-}$ channels, and large conductance Ca$^{2+}$-dependent K$^+$ channels (BK)~\citep{ghatta2006large}. \ac{CICR} facilitates Ca$^{2+}$ release from intracellular stores~\citep{bootman2002calcium}, while \ac{VOCC} mediate calcium entry during depolarization, crucial for neurotransmitter release and muscle contraction~\citep{sher1991physiopathology}. Other critical ion transporters include \ac{KIR}, which regulate the resting membrane potential~\citep{hibino2010inwardly}, the \ac{NBC} for pH and ion homeostasis~\citep{bernardo2006sodium}, and the \ac{KCC1} for cell volume regulation~\citep{russell2000sodium}. Neuronal communication relies on electrical potentials, with dendrites transmitting signals to the soma, where action potentials are generated~\citep{heinrich2014loffler,brandes2019physiologie}. Action potentials involve sequential phases: overcoming the threshold potential, rapid depolarization (upstroke and overshoot), repolarization, and post-hyperpolarization, mediated by Na$_v$-channels for depolarization and K$_v$-channels for repolarization~\citep{brandes2019physiologie}.
\\
\indent Axons and dendrites function as biological cables composed of capacitor-resistor elements, with distinct roles in neural signal processing. In dendrites, the membrane time constant and length constant govern the integration and conduction of synaptic potentials to the soma, while axons employ active mechanisms to propagate action potentials~\citep{brandes2019physiologie}. Unmyelinated axons typically conduct action potentials at approximately 1~m/s, with conduction speed increasing as axonal diameter grows, reducing internal resistance. Human unmyelinated nerve fibers generally have a diameter of around 1~$\mu$m, allowing continuous excitation propagation along the axon to the synapse~\citep{brandes2019physiologie}. At chemical synapses, electrical signals are converted into chemical reactions mediated by neuro- and gliotransmitters~\citep{heinrich2014loffler}. Synaptic transmission involves a sequence of events: nerve terminal depolarization, Ca$^{2+}$ influx via \ac{VOCC}, diffusion of Ca$^{2+}$ to vesicular active sites, neurotransmitter release into the synaptic cleft, receptor binding on the postsynaptic membrane, activation of ion channels or second messenger systems, and either depolarization or hyperpolarization of the postsynaptic membrane~\citep{brandes2019physiologie}.
\subsection{Physiology of Vasoconstriction, Autoregulation and Bioenergetics}
Understanding the physiology of vasoconstriction, autoregulation, and bioenergetics is essential for accurately modeling the dynamic interactions within the \ac{NGVU}. These processes govern the regulation of cerebral blood flow, vascular tone, and intracellular energy metabolism, forming the basis for predictive models of brain function. By incorporating these mechanisms, models can capture the physiological responses to neural activity, systemic changes, and metabolic demands, enabling deeper insights into neurovascular coupling and brain homeostasis.

\ac{SMCs} in the blood vessel wall form a syncytium and rely on post-translational modifications for contraction due to the absence of intrinsic myosin ATPase activity. Contraction is regulated by \ac{MLCK}, a Ca$^{2+}$/\ac{CaM}-dependent kinase activated by free Ca$^{2+}$, which phosphorylates myosin, enabling actin-myosin interaction and force generation~\citep{ito1990phosphorylation, dormanns2015neurovascular, zilles2011anatomie}. Caldesmon, bound to actin-tropomyosin, is displaced upon Ca$^{2+}$-\ac{CaM} binding, facilitating this interaction. \ac{MLCP} counteracts \ac{MLCK} activity, its regulation mediated by Rho-kinase inhibition or stimulation via \ac{NO} through \ac{cGMP} and \ac{cAMP} production~\citep{PAPPANO2013171}. \ac{SMCs} integrate inputs from neurons, astrocytes, and \ac{ECs}, primarily via Ca$^{2+}$ transport and voltage coupling. Their contraction or relaxation modulates vessel radius, driving vasoconstriction or dilation and playing a critical role in neurovascular regulation within the NGVU~\citep{dormanns2015neurovascular, PAPPANO2013171}.

The cardiovascular system, consisting of the heart and blood vessels, facilitates the transport and exchange of \ac{O$_2$}, \ac{CO$_2$}, nutrients, electrolytes, and hormones~\citep{zilles2011anatomie}. Arteries carry blood away from the heart, while veins return it, with arterioles, capillaries, and venules connecting these pathways. The aorta branches into progressively smaller arteries, forming a vascular tree where vessel diameters decrease while cross-sectional area increases. Vessel walls have three layers: the tunica intima, lined with \ac{ECs} for adhesion and coagulation; the tunica media, containing \ac{SMCs} and elastic fibers to regulate diameter; and the tunica externa, made of connective tissue~\citep{zilles2011anatomie}. Arteries can measure up to $\SI{3}{\centi\metre}$, arterioles are under $\SI{30}{\micro\metre}$, and capillaries are $5$--$\SI{8}{\micro\metre}$ wide, optimizing exchange due to their large cross-sectional area and a flow rate of $\SI{0.5}{\milli\metre\per\second}$~\citep{zilles2011anatomie, xue2022quantification}. Four major arteries, including the internal carotid and vertebral arteries, supply the brain. Capillaries, spanning $\SI{644}{\kilo\metre}$ with a surface area of $\SI{20}{\metre\squared}$, consist of \ac{ECs}, a basal lamina, and pericytes. Smooth muscle layers in larger arteries reduce to a single layer in capillaries, where pericytes replace \ac{SMCs}, separated from astrocytes by the perivascular space~\citep{schaeffer2021revisiting, begley2003structural, zlokovic2008blood}.

Autoregulation maintains nearly constant \ac{CBF} within a range of \SIrange{50}{160}{mmHg} by adjusting vascular tone to ensure oxygen and nutrient delivery~\citep{elting2020assessment,duchemin2012complex, iadecola1993regulation}. \ac{CBF} is driven by \ac{CPP}, defined as the difference between \ac{MAP} and \ac{ICP}, and cerebral vascular resistance (CVR)~\citep{presa2020vasculo}. \ac{CPP} typically ranges from \SIrange{60}{80}{mmHg}, while normal \ac{ICP} is \SIrange{5}{10}{mmHg} and has less impact on \ac{CPP} than \ac{MAP}~\citep{presa2020vasculo}. \ac{MAP}, averaging blood pressure over one cardiac cycle, normally ranges from \SIrange{70}{100}{mmHg} but fluctuates with activity, stress, or other systemic factors~\citep{mount2023cerebral}. Autoregulation alters vessel diameter through vasodilation (increased diameter) or vasoconstriction (decreased diameter)~\citep{peppiatt2006bidirectional, phillips2016neurovascular}. However, its range is limited to $\pm\SI{20}{mmHg}$ from baseline with a latency of \SI{60}{s}~\citep{claassen2021regulation}. Sudden or extreme \ac{AP} changes induce passive \ac{CBF} variations. Elevated AP increases intracellular Ca$^{2+}$ in \ac{SMCs}, activating \ac{MLCK} and triggering actin-myosin crosslinking, resulting in smooth muscle contraction and vasoconstriction~\citep{schaeffer2021revisiting, phillips2016neurovascular}. Systemic factors such as posture, glucose levels, hormones, hematocrit, blood viscosity, and cold stress also influence \ac{CBF}, engaging the cerebrovascular tree with segmental specificity~\citep{claassen2021regulation, barnes2021integrative}.

Intracellular nutrient metabolism is fundamental for brain function, driven by biochemical pathways that depend on the delivery of glucose and oxygen, regulated by neuro-glial activity. Within cells, mitochondria play a central role in energy production through enzymatic reactions, essential for eukaryotic cells such as neurons, astrocytes, \ac{ECs}, and \ac{SMCs}~\citep{cortassa2019control}. Metabolic pathways include glycolysis, glycogenolysis, the pentose phosphate and polyol pathways, $\beta$-oxidation, the \ac{TCA} cycle, \ac{ATP} synthesis, ionic and redox-energy transducing processes, and \ac{ROS} generation and scavenging~\citep{cortassa2019control}. \ac{ATP} is synthesized via glucose catabolism through glycolysis, the \ac{TCA} cycle, and oxidative phosphorylation, where \ac{NADH}-driven proton gradients power \ac{ATP} generation~\citep{senior2002molecular}. Glycolysis produces \ac{ATP} both anaerobically and aerobically, generating pyruvate for the \ac{TCA} cycle and intermediates for anabolic pathways~\citep{chandel2021glycolysis}. Glycogenolysis breaks glycogen into glucose-1-phosphate and glucose, providing a key glycogen utilization pathway~\citep{panja2013closed}. The pentose phosphate and polyol pathways support biosynthetic metabolism and redox homeostasis~\citep{stincone2015return}. The \ac{ATP} cycle, involving eight enzymes, links pyruvate and malate oxidation to CO$_2$ production and NADH generation, sustaining mitochondrial respiration and metabolic functions~\citep{fernie2004respiratory, fernie2009malate}. $\beta$-oxidation of \ac{PUFAs} balances energy demand, competing with glucose as a primary oxidative substrate~\citep{ghisla2004beta}. Ionic and redox-energy transducing processes couple biochemical systems through electronic energy transitions~\citep{losada1983energy}. \ac{ROS} generation and scavenging involve oxygen radicals and oxidizing agents, influencing metabolism, mitochondrial electron transport, signal transduction, and gene expression~\citep{bayr2005reactive}.

\section{Neuro-Glial-Vascular Unit Model}\label{sec:NGVU_Model}
\begin{figure}[h!]
    \centering
    \includegraphics[width=1.0\linewidth]{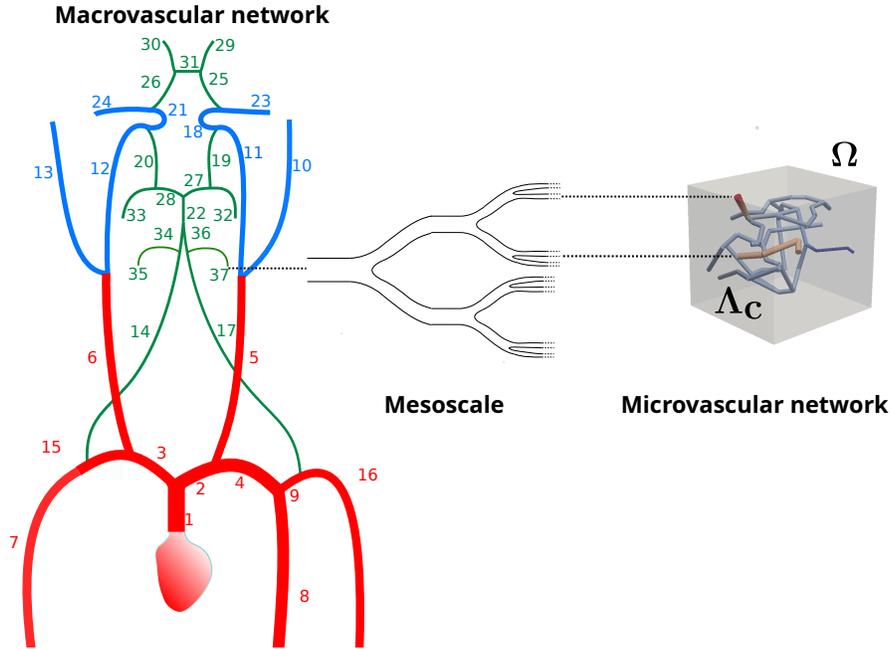}
    \caption{\label{fig:Networks} This figure gives an overview on the vascular systems that are considered for simulating blood flow from the heart to the neuro-glial-vascular unit. The macrovascular network and mesoscaled network are part of the arterial tree, while the microvascular network has to be assigned to the capillaries. By means of the macrovascular network, we determine the amount of blood volume transported towards the head in particular through the vertebral arteries. The mesoscaled network connects one of the vertebral arteries with the capillary network $\Lambda_c$ supplying our neuro-glial unit contained in $\Omega$.}
\end{figure}
\subsection{Blood flow model}
In order to determine the blood supply of a \ac{NGVU}, we consider a capillary network contained in a block measuring $150 \times 160 \times 140~\mu\text{m}$. It is based on a three-dimensional vascular network reconstructed from \ac{SEM} data of the rat brain. The vascular structure, originally presented by~\citep{motti1986terminal}, highlights the terminal vascular bed in the superficial cortex. The reconstruction and its application to oxygen transport simulations were detailed by~\citep{secomb2000theoretical}. This dataset, denoted by \emph{brain99}, provides geometric and flow data for a vascular block $\Omega \subset \mathbb{R}^3$ measuring $150 \times 160 \times 140~\mu\text{m}$, including information on vessel radii, lengths, and Cartesian coordinates.

The data set \emph{brain99} is utilized as a representative microcirculation block for the \ac{DVC}. At the inlets of the capillary vessels with the two largest inlets, coupling between macro- and microcirculation is applied, ensuring a physiologically consistent transition of flow and pressure between arterial and capillary blood vessel networks. For this purpose, we prescribe at the respective inlets flow rates derived from a model for blood flow in a network composed of larger arteries. The network, which is considered in this work can be seen in Figure~\ref{fig:Networks}. It consists of the largest arteries attached to the heart, as well as the brachial and carotid arteries, and the circle of Willis. This network is enhanced by two arteries that branch out from the vertebral arteries (Vessel $35$ and $37$ in Figure~\ref{fig:Networks}). Moreover, the vertebral arteries are split into two parts, where the distal parts are considered as new vessels. The four additional vessels are labeled with the indices $34$, $35$, $36$ and $37$. In the remainder of this work, this network is referred to as a macrovascular network. We assume that $\Omega$ is supplied by Vessel $37$. In order to connect $\Omega$, we use an averaged flow rate at the outlet of Vessel $37$ as input for a simple model representing the vessel tree linking Vessel $37$ and the capillary network $\Lambda_c$ contained in $\Omega$. This vessel tree comprises a part of the mesoscale within the vascular tree. An an output of the mesoscale model, we obtain two flow rates that are prescribed at the inlets of the vessels with the two largest radii (see Figure~\ref{fig:Networks}). All in all we obtain a multi-scale model for a flow path starting at the inlet of the aorta and ending in the \ac{NGVU} contained in $\Omega$.

In the following, we describe the different components of our model: In a first subsection, a dimension-reduced model for flow in the macrovascular network is presented. Next, a flow model within the microvascluar network is described. Finally, a strategy for coupling both models via a model for the mesoscale is discussed.

\subsubsection{Blood flow model for the macrovascular network}
\label{sec:macrocirculation}\leavevmode\\
For the macrovascular network we use the one-dimensional model (1D) described in \cite{vcanic2003mathematical,hughes1973one} and \cite[Chapter 2]{d2007multiscale} on each vessel. 
Hence, we determine  the vessel area $A\;\left[ \unit{cm}^2 \right]$, and flow rate $Q\;\left[ \unit{cm}^3\unit{s}^{-1}\right]$ on the center line $\Lambda$ of a blood vessel with length $L\;\left[ \unit{cm} \right]$. $\Lambda$ is parameterized by a variable $z\in \left[0,L \right]$. $A$ and $Q$ are solutions of a non-linear hyperbolic system of equations given by \cite[Section 3.2]{koppl2023dimension}: 
\begin{align}
\label{eq:masscon}  &\partial_t A + \partial_z Q = 0, \;z \in \left(0,L \right),\;t>0, \\
\label{eq:momentumcon} &\partial_t Q + \partial_z\left( \frac{Q^2}{A} \right) + \rho^{-1}A \,\partial_z P = - 2 \cdot \left( \bar\gamma+2 \right) \cdot \frac{\mu\left( 2 \cdot R \right)}{\rho} \cdot \frac{Q}{A},\;z \in \left(0,L \right),\;t>0,
\end{align}
where $t\;\left[\unit{s}\right]$ is the time variable and $\rho = 1.028 \cdot 10^{-3} \;\unit{g}\,\unit{cm}^{-3}$ represents the fluid density. $\mu\;\left[\unit{Pa\;s}\right]$ is a function describing the viscosity depending on the vessel radius $R\;\left[\unit{cm}\right]$. The formula for this function can be found in \cite{pries1996biophysical}. 
$\bar\gamma = 9$ is a shape parameter for the flow profile~(\cite[Cp. 2 \& Sec.~6.1]{d2007multiscale}). The pressure is related to the vessels by a fluid-structure interaction model, which is known as the Young-Laplace equation~\cite{olufsen1999structured,toro2016brain}:
$$ 
P = G_{0} \left( \sqrt{\frac{A}{A_{0}}} - 1 \right).
$$
The parameter $G_0$ is given by:
$$
G_{0} = \frac{\sqrt{\pi} \cdot h_{0} \cdot E}{\left(1-\nu^2\right) \cdot \sqrt{A_{0}}}.
$$
$h_0\,\left[\unit{cm}\right]$ is the wall thickness, $E\,\left[\unit{Pa}\right]$ the elastic modulus, $\nu = \tfrac {1}{2}$ is the Poisson ratio and the vessel area for $P=0$ is denoted by $A_0\,\left[\unit{cm}^2\right]$. The radius of $A_0$ is given by $R_0\,\left[\unit{cm}\right]$. 

A table containing the values for $L$, $h_0$, $E$ and $R_0$ can be found in \cite[Table 1]{alastruey2007modelling}. In this table one can find the data for Vessel $1$-$33$ apart from the lengths of Vessel $14$ and $17$. The missing data for Vessel $34$ to $37$ as well as Vessel $14$ and $17$ can be listed in Table \ref{tab:Param3437}.
\begin{table}[h!]
    \centering
    \begin{tabular}{|c|c|c|c|c|}
      \hline
      \hline
      Vessel   &  $L\,\left[\unit{cm}\right]$  &  $R\,\left[\unit{cm}\right]$ &  $h_0\,\left[\unit{cm}\right]$ & $E\,\left[10^6 \cdot \unit{Pa}\right]$ \\
      \hline
       $14$    & $9.866$  & $0.136$ & $0.034$ & $0.8$ \\
      \hline
       $17$    & $9.866$  & $0.136$ & $0.034$ & $0.8$ \\
      \hline
       $34$    & $4.930$  & $0.136$ & $0.034$ & $0.8$ \\
      \hline
       $35$    & $6.000$  & $0.080$ & $0.034$ & $0.8$ \\
      \hline
       $36$    & $4.930$  & $0.136$ & $0.034$ & $0.8$ \\
      \hline
       $37$    & $6.000$  & $0.080$ & $0.034$ & $0.8$ \\
      \hline
      \hline
    \end{tabular}
    \vspace{0.2cm}
    \caption{Parameters for Vessel $14$ and $17$ as well as Vessel $34$ to $37$.}
    \label{tab:Param3437}
\end{table}
\ \\
At the bifurcations of the network, we enforce continuity of the total pressure and flow conservation. The heart beats are emulated by describing the flow rate at the inlet of the aorta (Vessel $1$) by a half-sine wave. We denote this flow rate by $Q_{in}\;\left[ \unit{cm}^3\unit{s}^{-1}\right]$. It is given by the $T$-periodic extension of 
$$
Q_{in}\left(t\right) = 
\begin{cases}
Q_{\textmd{max}} \sin \left( \frac{\pi \cdot  t}{0.3 \cdot T} \right), &\text{ for } 0 \leq t \leq 0.3 \cdot T \\
0, &\text{ for } 0.3 \cdot T < t \leq T.
\end{cases}
$$
In case of our simulations, we set $T = 1.0\;\unit{s}$ and $Q_{\textmd{max}}=485.0\;\unit{cm}^3\unit{s}^{-1}$. The influence of the omitted vasculature beyond the outlets is accounted for by a zero-dimensional (0D) Windkessel model~\cite{fritz20221d}. To calibrate the Windkessel parameters, a method described in \cite[Section 3.6.2]{koppl2023dimension} is used. All in all, this results in a 1D-0D coupled model. 
\subsubsection{Blood flow model for the microvascular network}
\label{sec:microcirculation}\leavevmode\\
Modeling blood flow in a capillary network, we use a stationary description relating the 1D pressures $p_1$ on the capillary centerlines $\Lambda_c$ with the 3D pressures $p_3$ in the surrounding space $\Omega$:
\begin{align}
	- R^2 \pi  \;\partial_{s} ( K_{1} \; \partial_{s} p_1 )   + 2 \pi R L_{p}( p_1 - \Pi_{\Omega}{p}_3) &= 0
   &\textmd{on } \Lambda_c,
 \\
   -\nabla \cdot ( K_3 \nabla p_3 ) + L_p ( p_3 - \Pi_\Lambda p_1 ) \delta_\Gamma &= 0
   &\textmd{on } \Omega,
\end{align}
where $\Pi_\Lambda: \Lambda_c \to \Gamma $ extends functions on the center line to functions on the cylinder boundary, $\Pi_{\Omega}: \Omega \to \Lambda$ averages functions on cylinder cross sections, and $\delta_\Gamma$ the Dirac distribution on the cylindrical vessel walls. By this, we obtain a 3D-1D coupled model. $R$ describes the radius of the vessels and $L_p$ is the permeability of the vessel walls. Moreover $K_1 = \nicefrac{R^2}{8 \mu\left(2 \cdot R \right)}$ and $K_3 $ are the permeabilities of the vessel and the medium contained in $\Omega$ (see~\cite{koppl20203d} for details). As in Section \ref{sec:macrocirculation}, the viscosity function $\mu$ is taken from~\cite{pries1996biophysical}. This flow model yields flow velocities in the capillary network $\Lambda_c$ and $\Omega$, which are denoted by $v_c$ and $v_\Omega$. The velocities are computed as follows:
$$
v_c = -K_1 \partial_{s} p_1 \text{ and } v_\Omega = -K_3\nabla p_3.
$$
Using these velocities, convection diffusion equations are formulate to model the transport of oxygen, nutrients and other substances within $\Lambda_c$ and $\Omega$. As in case of the flow model there is a 1D PDE for the transport in the capillary network and a 3D PDE for the corresponding processes in the surrounding medium. For further information, we refer to~\cite{koppl20203d}. At the boundary of $\Omega$ we set homogeneous Neumann boundary conditions. The boundary conditions for $\Lambda_c$ are discussed in the next subsection.
\subsubsection{Coupling of macro- and microcirculation}\leavevmode\\
To couple the macrocirculation with the microcirculation, we assume that the inlets in our capillary bed are connected to the outlet of Vessel $37$ by a surrogate model for the arterial tree between the capillary network and the outlet of Vessel $37$ (see Figure \ref{fig:Networks}). As in~\cite{fritz20221d}, it is assumed that these arterial trees are symmetric. This means that every vessel bifurcates into two child vessels following Murray's law~\cite{murray1926physiological,murray1926physiological2}, and additionally, both child vessels have the same radii. Vessel radii between two arbitrary bifurcation levels $l$ and $l+1$ are thus related by
$$
r_l = 2^{1/\gamma} r_{l+1},
$$
where $\gamma \in \left[2,3.5\right]$. If $n$ bifurcations are given, we $r_0$ and the final radius $r_n$ are related to each other as follows: 
$$
r_0 = 2^{n/\gamma} r_{n}.
$$
The number of levels in our arterial tree is thus given by 
$$
n = \log_2( r_0/r_n)^{\gamma}.
$$
Assuming that the flow $Q_l$ at level $l$ is evenly divided at each bifurcation yields the relationship $Q_l = 2 Q_{l+1}$. This implies:
$$
Q_0 = 2^n Q_n = \left( \frac{r_0}{r_n} \right)^\gamma Q_n.
$$
Coupling the macro- and microcirculation, we replace $r_0$ by the initial radius of Vessel $37$: $r_0 = R_{0,37}$, while $r_n$ is the radius of a capillary containing an inlet. In case of our capillary network $\Lambda_c$, we choose the vessels at the boundary with the two largest radii. The corresponding radii are denoted by $r_{c,i},\;i\in \left\{1,2\right\}$. To compute $Q_0$, we simulate blood flow within the macrovascular network starting with $A=A_0$ and $Q=0$ for each vessel. The simulation is stopped after the pressure and flow rate curves become periodic for each heart beat. If this is the case, we report the flow rate curve $Q_{37}$ at the outlet of Vessel $37$ for one heart beat i.e. for a time interval 
$\left[t,t+T\right]$ and compute the average of $Q_{37}$. Finally, we equate $Q_0$ and this average:
$$
Q_0 = \frac{1}{T} \int_{t}^{t+T}Q\left(t\right)\;dt.
$$
By this, the flow rates $Q_{r_{c,i}}$ are given by:
$$
Q_{r_{c,i}} = Q_0 \cdot \left( \frac{r_{c,i}}{R_{0,37}} \right)^\gamma
$$
To ensure mass conservation, we distribute at the outlets of $\Lambda_c$ the flow rate 
$$
Q_{in,c} = Q_{r_{c,1}} + Q_{r_{c,2}}
$$
entering the network according to their section areas. Since we have on both inlets and outlets Neumann boundary conditions, we have to fix the pressure at some place of the network such that an average pressure of about $30\;\unit{mmHg}$ in $\Lambda_c$ results. Otherwise the flow problem for $\Lambda_c$ would be ill-posed.
\subsection{Quadripartite Synapse Model}
The quadripartite synapse model of the \ac{DVC} incorporates the Glu-, GABA-, neurons and astrocytes. The neurotransmission involves Glu, GABA, AMPA, NMDA, GABAa and GABAb type receptors, as described in~\citep{brazhe2023astrocytes, tewari2012mathematical}. The model encompasses a pre-synaptic bouton, a post-synaptic dendritic spine-head, \ac{SMCs}, and a perisynaptic astrocytes that governs Ca$^{2+}$ dynamics within the synaptic bouton. The calcium dynamics subsequently influence Glu release in the \ac{SMCs}. The model further accounts for Glu concentration within the \ac{SMCs}, incorporating the process of Glu reuptake by astrocytes. We further incorporated GABA neurons which have similar dynamics to the Glu-neurons. In addition, NMDA, GABAa and GABAb receptors with similar dynamics as AMPA receptor are considered.

Neuronal activities have the ability to induce elevations in Ca$^{2+}$ within astrocytes~\citep{fellin2009communication, porter1996hippocampal}, which in turn increases the concentration of Ca$^{2+}$ in neighboring glial cells, including astrocytes, that express a wide range of receptors~\citep{newman2003new}. The activation of these receptors leads to an increase in astrocytic [Ca$^{2+}$] and the subsequent release of various transmitters such as Glu, GABA, d-serine, and ATP. The depletion of the releasable vesicle pool due to recent neurotransmitter release in response to pre-synaptic action potentials results in short-term synaptic depression~\citep{wu1999reduced,brazhe2023astrocytes}. The proposed model is inspired by the experimental work conducted in~\citep{perea2007astrocytes}, who utilized hippocampal slice preparations from immature Wistar rats. The computational framework presented here integrates various detailed biophysical models to emphasize distinct facets of neurons-astrocytes signaling. 

\subsubsection{Pre-synaptic Neuronal Action Potential and Bouton Calcium Dynamics}\leavevmode\\
The generation of pre-synaptic action potential trains is achieved using the Hodgkin-Huxley formalism~\citep{hodgkin1952quantitative}. Elevation of Ca$^{2+}$ concentration within the pre-synaptic bouton is modeled by incorporating both rapid influx, utilizing single protein properties~\citep{erler2004quantitative}, and slower dynamics through a modified Li-Rinzel model~\citep{li1994equations}. The initiation of the action potential occurs at the axon hillock of the pre-synaptic neurons. The generation of pre-synaptic action potentials in computational models has traditionally been accomplished using the Hodgkin-Huxley framework~\citep{nadkarni2003spontaneous,volman2007astrocyte}. Given that the details of pre-synaptic action potential generation are not the primary focus of the present study, we adopt the Hodgkin-Huxley model to simulate regular spiking and burst activity for simplicity, given by
\begin{align}
    \label{eqn:pre_synaptic_ne}
    \begin{split}
        C \frac{\mathrm{d}V_{pre}}{\mathrm{d}t} =& I_{app} - g_K n^4 (V_{pre}-V_K) - g_{Na}m^3  h (V_{pre}-V_{Na})-g_L (V_{pre}-V_L), \\
        \frac{\mathrm{d}x}{\mathrm{d}t} =& \alpha_x (1-x)-\beta_x x, \qquad x\in\{n,m,h\},  
    \end{split}
\end{align}
where $V_{pre}$ denotes the pre-synaptic membrane potential, and $I_{app}$ represents the applied current density. The conductances for potassium, sodium, and leak channels are given by $g_K$, $g_{Na}$, and $g_L$ respectively. Correspondingly, $V_K$, $V_{Na}$, and $V_L$ are the reversal potentials for potassium, sodium, and leak channels. The variables $m$  denotes the sodium activation, $h$ is the sodium inactivation, and $n$ is defined as the potassium activation. The expressions for $\alpha_x$ and $\beta_x$ for $x=(m,h,n)$ are given by
\begin{alignat}{2}
        &\alpha_n = \frac{0.01 (-V_{pre}-60)}{\exp \left (\frac{-V_{pre}-60}{10} \right)-1}, \quad &&\beta_n = 0.125 \exp \left(\frac{-V_{pre}-70}{80}\right), \\ 
        &\alpha_m = \frac{0.1 (-V_{pre}-45)}{\exp\left(\frac{-V_{pre}-45}{10}\right)-1},
        \quad &&\beta_m = 4\exp \left(\frac{-V_{pre}-70}{18} \right), \\
        &\alpha_h = 0.07 \exp \left(\frac{-V_{pre}-70}{20}\right), 
        \quad &&\beta_h =\frac{1}{\exp\left(\frac{-V_{pre}-40}{10}\right)+1}.
\end{alignat}
The action potential initiated at the axon hillock of the pre-synaptic neurons propagates along the axon to the terminal without degradation, causing an elevation in cytosolic [Ca$^{2+}$]. This increase in intracellular [Ca$^{2+}$] is composed of two parts: [Ca$^{2+}$] resulting from the action potential, denoted as $c_{fast}$, and [Ca$^{2+}$] originating from intracellular stores, denoted as $c_{slow}$, based on the kinetics speed. Thus, the total intracellular [Ca$^{2+}$], represented as $c_i$, is defined as
\begin{equation}
    c_i = c_{fast}+c_{slow} \rightarrow \frac{\mathrm{d}c_i}{\mathrm{d}t} = \frac{\mathrm{d}c_{fast}}{\mathrm{d}t}+\frac{\mathrm{d}c_{slow}}{\mathrm{d}t}.
\end{equation}
The significant impact of rapid Ca$^{2+}$ kinetics on neurotransmitter release is well documented~\citep{bollmann2000calcium}. For simplicity, the Ca$^{2+}$ influx through the plasma membrane is modeled via N-type channels~\citep{weber2010n}. Immature cells, as described in~\citep{perea2007astrocytes}, are assumed in this study. The equation governing [Ca$^{2+}$] resulting from the action potential follows a straightforward construction and degradation framework~\citep{keener2009mathematical}, which can be expressed as 
\begin{equation}
\label{eqn:fast_c}
    \frac{\mathrm{d}c_{fast}}{\mathrm{d}t} = \underbrace{-\frac{I_{Ca}A_{btn}}{z_{Ca}FV_{btn}}+J_{PM,leak}}_{construction} \underbrace{- \frac{I_{PM,Ca}A_{btn}}{z_{Ca}FV_{btn}}}_{destruction}, 
\end{equation}
where $I_{Ca}$ represents the Ca$^{2+}$ current through an N-type channel, $A_{btn}$ is the bouton's surface area, $z_{Ca}$ is the valence of the Ca$^{2+}$ ion, $F$ is Faraday's constant, and $V_{btn}$ is the bouton's volume. Additionally, $I_{PM,Ca}$ denotes the current attributed to the electrogenic plasma membrane Ca$^{2+}$ ATPase, which expels excess Ca$^{2+}$ from the cell~\citep{jensen2007presynaptic}. The function of this pump is described using a standard Michaelis-Menten type formalism~\citep{erler2004quantitative}. Furthermore, $J_{PM,leak}$ accounts for the positive leak of Ca$^{2+}$ from the extracellular space into the bouton, ensuring that the Michaelis-Menten pump does not reduce cytosolic Ca$^{2+}$ to zero~\citep{blackwell2005modeling}. The Ca$^{2+}$ current through the N-type channel is modeled at the single protein level~\citep{erler2004quantitative} as
\begin{equation}
    I_{Ca} = \rho_{Ca} m^2_{Ca} \underbrace{g_{Ca}(V_{pre}(t)-V_{Ca})}_{single~open~channel},
\end{equation}
where $\rho_{Ca}$ represents the density of N-type channel proteins, determining the number of Ca$^{2+}$ channels present on the bouton membrane. The parameter $g_{Ca}$ denotes the conductance of a single N-type channel, while $V_{Ca}$ is the reversal potential for the Ca$^{2+}$ ion, given by the Nernst equation~\citep{keener2009mathematical} as 
\begin{equation}
\label{VCa}
    V_{Ca} = \frac{RT}{z_{Ca}F} \ln{\left(\frac{c_{ext}}{c_i^{rest}}\right)},
\end{equation}
where $R$ denotes the universal gas constant, $T$ is the absolute temperature, $c_{ext}$ represents the extracellular Ca$^{2+}$ concentration, and $c_i^{rest}$ signifies the resting total intracellular [Ca$^{2+}$]. It is assumed that each N-type channel is composed of two gates, with $m_{Ca}$ indicating the probability of a single gate being open. Therefore, the probability of a single N-type channel being open, which requires both gates to be open, is given by $m^2_{Ca}$. The time evolution of this single-channel open probability follows a Hodgkin-Huxley type model. The steady-state value $m^\infty_{Ca}$, as determined according to~\citep{ishikawa2005presynaptic}, fits the Boltzmann function to the whole-cell current of an N-type channel. The variable $m_{Ca}$ approaches its asymptotic value $m^\infty_{Ca}$ with a time constant $\tau_{m_{Ca}}$ expressed by means of an ordinary differential equation as
\begin{equation}
    \frac{\mathrm{d}m_{Ca}}{\mathrm{d}t} = \frac{(m^\infty_{Ca}-m_{Ca})}{\tau_{m_{Ca}}}.   
\end{equation}
Additional mathematical expressions for the parameters utilized in~\eqref{eqn:fast_c} are given by
\begin{align}
    I_{PM,Ca} &= v_{PM,Ca} \frac{c_i^2}{c_i^2 + K^2_{PM,Ca}}, \\
    J_{PM,leak} &= v_{leak} (c_{ext}-c_i), \\
    m^{\infty}_{Ca} &= \frac{1}{1+\exp{((V_{m_{Ca}}-V_m)/k_{m_{Ca}})}},
\end{align}
where $v_{PMCa}$ denotes the maximum \ac{PMCa} current density, determined through computer simulation, so that $c_i$ is maintained at its resting concentration. 
The second component of bouton Ca$^{2+}$, denoted as $c_{slow}$, is recognized for its essential role in short-term plasticity~\citep{emptage2001calcium}. The release of Ca$^{2+}$ from the \ac{ER} is primarily regulated by two types of receptors or Ca$^{2+}$ channels: the IP$_3$ receptor and the ryanodine receptor~\citep{sneyd2005models}. In this context, the flow is considered to occur solely through the IP$_3$ receptor. The IP$_3$ required for the release of Ca$^{2+}$ from the \ac{ER} is generated when Glu binds to metabotropic Glu-receptors, activating a G-protein-linked pathway to phospholipase C, which then cleaves phosphatidylinositol (4,5)-bisphosphate (PIP$_2$) to produce IP$_3$ and diacylglycerol. For modeling this slower Ca$^{2+}$ signaling process, an adapted Li-Rinzel model is employed (\emph{cf}.~\citep{tsodyks1997neural}). Originally, the model assumes the total intracellular concentration $c_0$, is conserved and determines the \ac{ER} Ca$^{2+}$ concentration, $c_{ER}$, following the relation
\begin{equation}
\label{eqn:c_er}
    c_{ER} = \frac{c_0-c_i}{c_1}.
\end{equation}
Due to the presence of membrane fluxes, specifically $I_{Ca}$ and $I_{PMCa}$, the assumptions from the original Li-Rinzel model are not applicable in the current model. Moreover, the Li-Rinzel model also incorporates intracellular IP$_3$ concentration as a control parameter. To address these discrepancies, two additional equations governing the \ac{ER} concentrations [Ca$^{2+}$] and [IP$_3$] have been incorporated into the Li-Rinzel model. The IP$_3$ production term has been made dependent on Glu to examine the effect of astrocytes Ca$^{2+}$ on $c_i$~(\emph{cf}.~\citep{nadkarni2007modeling}). The mathematical model governing the $c_{slow}$ dynamics is thus given by
\begin{align}
\label{eqn:pre_synaptic_ne0}
        \frac{\mathrm{d}c_{slow}}{\mathrm{d}t} &= -J_{chan}-J_{ER,pump}-J_{ER,leak}, \\
    \frac{\mathrm{d}c_{ER}}{\mathrm{d}t} &= - \frac{1}{c_1} \frac{\mathrm{d}c_{slow}}{\mathrm{d}t}, \\
    \frac{\mathrm{d}p}{\mathrm{d}t} &= v_g \frac{g_a^{0.3}}{k^{0.3}_g+g^{0.3}_a} - \tau_p (p-p_0), \\
    \frac{\mathrm{d}q}{\mathrm{d}t} &= \alpha_q (1-q) -\beta_q q.
\end{align}
Here, $J_{chan}$ represents the Ca$^{2+}$ flux from the \ac{ER} to the intracellular space through IP$_3$ receptors, while $J_{ER,pump}$ denotes the Ca$^{2+}$ flux being pumped from the intracellular space back into the \ac{ER}. The variable $c_{ER}$ indicates the \ac{ER} Ca$^{2+}$ concentration, $c_1$ is the ratio of \ac{ER} volume to bouton volume, $p$ represents the intracellular IP$_3$ concentration, $g_a$ signifies the Glu in the extra-\ac{SMCs}, and $q$ denotes the fraction of activated IP$_3$ receptors. These fluxes are described by the following expressions:
\begin{align}
    J_{chan} &= c_1 v_1 m^3_\infty n^3_\infty q^3 (c_i-c_{ER}), \\ 
    J_{ER,pump} &= \frac{v_3 c_i^2}{k^2_3 +c_i^2}, \\
    J_{ER,leak} &= c_1 v_2 (c_i -c_{ER}),
\end{align}
with $m_\infty = \tfrac{p}{p+d_1}$, $n_\infty = \tfrac{c_i}{c_i + d_5}$, $\alpha_q = a_2 d_2 \tfrac{p+d_1}{p+d_3}$ and $\beta_q=a_2c_i$. The values of $v_1$, $v_2$, and $v_3$ are iteratively determined to ensure that Ca$^{2+}$ homeostasis is maintained within the cell and its organelles.

\subsubsection{Glu- and GABA-Neuron Dynamics in the Bouton and Smooth Muscle Cell}\label{sec:glu_gaba}\leavevmode\\
The dynamics of Glu-neuron interactions are analogous to those observed for GABA-neurons interactions, where Glu has an excitatory effect and GABA is an inhibitor~\citep{petroff2002book}. In the following sections, the mathematical model will be presented, focusing on the Glu-neurons system and its dynamics within the \ac{SMCs} and the extra-\ac{SMCs}. In the quadripartite synapse model, both Glu and GABA are incorporated, each with distinct kinetic constants.
Glu release within the \ac{SMCs} is represented as a two-step process: Ca$^{2+}$ binding to a synaptic vesicle sensor is described according to~\citep{bollmann2000calcium}, while synaptic vesicle fusion and recycling follow the framework outlined in~\citep{tsodyks1997neural}. The elevation of extra-synaptic Glu is also approached as a two-step process: the probability of synaptic-like micro-vesicle release is fitted using a modified model of~\citep{bertram1996single}, and synaptic-like micro-vesicle fusion and recycling are again modeled following the Tsodyks-Markram model~\citep{tsodyks1997neural}, incorporating recent empirical data from~\citep{malarkey2011temporal}.

The action potential waveforms result in a temporary rise in intracellular [Ca$^{2+}$], which subsequently triggers neurotransmitter release~\citep{bollmann2000calcium,wang2009action}. Investigating the sensitivity of Ca$^{2+}$ sensors is challenging due to the small size of nerve terminals~\citep{wang2009action}. It is generally assumed that a Ca$^{2+}$ concentration of at least 100~$\mu$M is required to activate a \emph{low-affinity} Ca$^{2+}$ sensor~\citep{nadkarni2003spontaneous}. However, recent studies conducted in the giant calyx of Held terminal have demonstrated that an intracellular Ca$^{2+}$ concentration of approximately 10~$\mu$M is sufficient for Glu release~\citep{bollmann2000calcium}. The kinetic model describing Ca$^{2+}$ binding to the Ca$^{2+}$ sensor is expressed by 
\begin{equation}
\label{eqn:pre_synaptic_ne2}
     X \overset{5 \alpha c_i}{\underset {\beta}{\rightleftharpoons}} X(c_i)_1 \overset {4 \alpha c_i}{\underset{2 \beta}{\rightleftharpoons}} X(c_i)_2 \overset {3\alpha c_i} {\underset {3 \beta}{\rightleftharpoons}} X (c_i)_3 \overset{2 \alpha c_i}{\underset{4 \beta}{\rightleftharpoons}} X(c_i)_4 \overset{\alpha c_i}{\underset{5 \beta}{\rightleftharpoons}} X(c_i)_5 \overset{\gamma}{\underset{\delta}{\rightleftharpoons}}X(c_i)^*_5,
\end{equation}
where the constants $\alpha$ and $\beta$ represent the Ca$^{2+}$ association and dissociation rates, respectively, while $\gamma$ and $\delta$ are Ca$^{2+}$-independent isomerization constants. The variable $X$ denotes the Ca$^{2+}$ sensor of a synaptic vesicle without any bound Ca$^{2+}$ ions, with the index number indicating the number of bound Ca$^{2+}$ ions. The isomer $X(c_i)_5^*$ is the form of $X(c_i)_5$ that is prepared for Glu release. The fraction of docked vesicles ready for release, $f_r$, has been determined using dynamic Monte Carlo simulations~\citep{fall2002computational} based on the kinetics described in~\eqref{eqn:pre_synaptic_ne2}, which depends on the state $X(c_i)_5^*$. This approach is necessary because the number of synaptic vesicles with five Ca$^{2+}$ ions bound cannot be estimated by the average vesicle pool due to the limited number of vesicles. The release of neurotransmitters spontaneously depends on the pre-synaptic Ca$^{2+}$ concentration~\citep{bollmann2000calcium,emptage2001calcium}. The number of vesicles ready for spontaneous release, $p_r$, is assumed to follow a Poisson process~\citep{nadkarni2008astrocytes}, which is given by the rate
\begin{equation}
\label{eqn:pre_synaptic_ne3}
    \lambda(c_i) = a_3 \frac{1}{\left(1+\exp{\left(\frac{a_1-c_i}{a_2}\right)}\right)},
\end{equation}
that is iteratively adjusted to align with the frequency of spontaneous vesicle release in the presence of an astrocytes, typically ranging from $1$ to $3$ events per second~\citep{kang1998astrocyte}. The vesicle fusion and recycling process is described by the Tsodyks-Markram model~\citep{tsodyks1997neural}. To incorporate the dependency on $p_r$, a slight modification has been made to the Tsodyks-Markram model, yielding
\begin{align}
    \frac{\mathrm{d}R}{\mathrm{d}t} &= \frac{I}{\tau_{rec}} - f_r R, \\
    \frac{\mathrm{d}E}{\mathrm{d}t} &= - \frac{E}{\tau_{intact}}+f_r R, \\
    I &= 1 -R -E.
\end{align}
Here, $R$ represents the fraction of releasable vesicles within the bouton, $E$ denotes the fraction of effective vesicles in the \ac{SMCs}, and $I$ signifies the fraction of inactive vesicles undergoing recycling. The variable $f_r$ can take values of 0, 0.5, or 1, corresponding to the number of vesicles ready for release, which are according to 0, 1, or 2. These values are determined either by the stochastic simulation of the kinetic model in~\eqref{eqn:pre_synaptic_ne2} or by generating a Poisson random variable with the rate given by~\eqref{eqn:pre_synaptic_ne3}. The time constants for vesicle inactivation and recovery are denoted by $\tau_{inact}$ and $\tau_{rec}$, respectively. Once a vesicle is released, the vesicle release process remains inactive for a duration of $\SI{6.34}{\milli\second}$ in case of Glu and in case of GABA $\SI{7.2}{\milli\second}$~\citep{dobrunz1997very}.

Using Monte Carlo simulations of a central glutamatergic synapse, Franks et al.~\citep{franks2002monte} demonstrated that glutamatergic signaling is spatially independent at these synapses. The Glu concentration within a bouton vesicle has been estimated to be \SI{60}{\milli\M}~\citep{danbolt2001glutamate}. Given that $E$ represents the effective fraction of vesicles in the \ac{SMCs}, the Glu concentration in the \ac{SMCs} can be mathematically expressed as
\begin{equation}
\label{eqn:pre_synaptic_ne5}
    \frac{\mathrm{d}g}{\mathrm{d}t} = n_v g_v E - g_c g,
\end{equation}
where $g$ represents the Glu concentration in the \ac{SMCs}, $n_v$ is the number of docked vesicles, $g_v$ denotes the vesicular Glu concentration, and $g_c$ is the rate of Glu clearance~\citep{destexhe1998kinetic}. Using this straightforward dynamic model, it is possible to estimate a range for the Glu concentration in the \ac{SMCs} from $0.24$ to $\SI{11}{\nano\M}$~\citep{danbolt2001glutamate,franks2002monte}, as well as to capture the time course of Glu in the \ac{SMCs} over approximately $\SI{2}{\milli\second}$~\citep{franks2002monte,clements1996transmitter}.
The governing Glu dynamics in the extra-\ac{SMCs} is described by 
\begin{equation}
    \frac{\mathrm{d}g_a}{\mathrm{d}t} = n^v_a g_a^v E_a - g^c_a g_a, 
\end{equation}
where $g_a$ is the Glu concentration in the extra-\ac{SMCs}, $n^v_a$ represents the readily releasable pool of synaptic-like micro-vesicles, $g^v_a$ is the Glu concentration within each synaptic-like micro-vesicle and $g^c_a$ is the clearance rate of Glu from the \ac{SMCs} due to diffusion and/or re-uptake by astrocytes. Furthermore, it is evident that the formula is similar to Equation~(\ref{eqn:pre_synaptic_ne5}). In this case, Glu acts on extra-synaptic metabotropic Glu receptors located on the pre-synaptic bouton. This input is used for the IP$_3$ production term in Equation~(\ref{eqn:pre_synaptic_ne0}). Unlike the densely packed synaptic vesicles of neurons, the synaptic-like micro-vesicles of astrocytes are less densely packed \citep{bezzi2004astrocytes}. Therefore, it is assumed that each synaptic-like micro-vesicle contains 60~mM of Glu and 20~mM of GABA~\citep{montana2006vesicular}.

\subsubsection{Astrocyte Calcium Dynamics and Gliotransmitter Release}\leavevmode\\
The modulation of astrocytic Ca$^{2+}$ levels by Glu is described using an astrocytes-specific G-ChI model (\emph{cf}.~\citep{de2009glutamate}). This model is referred to as the G-ChI model, based on the dependent variables and the Glu concentration parameter utilized. Within this framework, $G$ signifies the Glu concentration in the \ac{SMCs}, $C$ is the astrocytic [Ca$^{2+}$], $h$ represents the gating variable of IP$_3$ receptors, and $I$ indicates astrocytic IP$_3$ concentration. The variable $g$ is treated as a dynamic variable as described in~\eqref{eqn:pre_synaptic_ne5}.
Within this model, the astrocytic Ca$^{2+}$ concentration, $c_a$, primarily depends on two parameters: the flux from the \ac{ER} into the cytosol and the maximum pumping capacity of the saroplasmatic reticulum ATPase pump. It is established that IP$_3$ receptors are organized in clusters within astrocytes~\citep{holtzclaw2002astrocytes}. The cluster size, $N_{IP3}$, is not precisely known and is assumed to be $20$ in this context (\emph{cf.}~\citep{shuai2002stochastic}). The stochastic Li-Rinzel model is employed here (\emph{cf}.~\citep{shuai2002stochastic}), which is expressed as
\begin{align}
\label{eqn:pre_synaptic_ne6}
    \begin{split}
        \frac{\mathrm{d}c_a}{\mathrm{d}t} & = (r_{Ca} m^3_\infty n^3_\infty h^3_a) (c_0 -(1+c_{1,a})c_a) - v_{ER} \frac{c_a^2}{c_a^2 +K^2_{ER}} \\
        &+ r_L (c_0 - (1+c_{1,a})c_a),
    \end{split} \\
\label{eqn:pre_synaptic_ne7}
    \begin{split}
        \frac{\mathrm{d}p_a}{\mathrm{d}t} &= v_\beta \text{Hill} \left(g^{0.9}, K_R \left(1+\frac{K_p} {K_R} \text{Hill} (C,K_\pi) \right) \right) + \frac{v_\delta}{1 + \frac{p_a}{k_\delta}} \text{Hill}(c_a^2, K_{PLC\delta}) \\
        &- v_{3K} \text{Hill}(c_a^4,K_D) \text{Hill} (p_a,K_3) -r_{5_{Pa}}p_a,
    \end{split} \\
\label{eqn:pre_synaptic_ne8}
    \begin{split}
        \frac{\mathrm{d}h_a}{\mathrm{d}t} &= \alpha_{h_a} (1-h_a) - \beta_{h_a}h_a+G_h (t).
    \end{split}
\end{align}
Equation~\eqref{eqn:pre_synaptic_ne6} consists of three terms on the right-hand side. The first one represents the Ca$^{2+}$ flux from the \ac{ER} into the intracellular space with $r_{Ca}$ denoting the maximal rate of the Ca$^{2+}$ flux from the IP$_3$ receptor, and $m^3_\infty n^3_\infty h^3_a$ together represent the opening probability of the IP$_3$ receptor cluster. The second term describes the rate at which Ca$^{2+}$ is removed from the intracellular space by the saroplasmatic reticulum ATPase pump with $v_{ER}$ representing the maximal rate of Ca$^{2+}$. The third term accounts for the leakage of Ca$^{2+}$ from the \ac{ER} into the intracellular space. Here, $r_L$ is the maximal rate of Ca$^{2+}$ leak from the \ac{ER}.
The terms involved depict similarity to the production terms of $c_{slow}$ in~\eqref{eqn:pre_synaptic_ne0}, with the main distinction being that this model operates under a closed-cell assumption. Given this, an expression like~\eqref{eqn:c_er} holds and can be represented using the astrocytic parameters as 
\begin{equation}
    c_{ER,a} = \frac{(c_0 -c_a)}{c_{1,a}}, \end{equation}
which has the advantage that the astrocytic Ca$^{2+}$ flux terms can be represented completely in terms of cell parameters. The second equation~\eqref{eqn:pre_synaptic_ne7} consists of four terms on the right-hand side. The first two terms represent the agonist-dependent and agonist-independent production of IP$_3$ and the last two incorporate the IP$_3$ degradation by IP$_3$-3K and IP$_3$-5P, respectively. In the last equation~\eqref{eqn:pre_synaptic_ne8}, $\alpha_{h_a}$ and $\beta_{h_a}$ denote the rates at which $h_a$ opens and closes, respectively, and $G_h (t)$ is a zero mean, uncorrelated, Gaussian white-noise term with a covariance function (\emph{cf}.~\citep{shuai2002stochastic}), given by
\begin{equation}
    \langle G_h (t), G_h (t') \rangle = \frac{\alpha_{h_a} (1-h_a) + \beta_{h_a} h_a}{N_{IP_3}} \delta (t-t'),
\end{equation}
where $\delta(t)$ is the Dirac-delta function, $t$ and $t'$ are distinct times and the fraction denotes the spectral density (\emph{cf}.~\citep{coffey2004langevin}). For astrocytic Ca$^{2+}$ oscillations, a model incorporating both amplitude and frequency modulation is used, as the coupling between IP$_3$ metabolism and \ac{CICR} does not support amplitude-only modulation~\citep{de2009glutamate}. The remaining expressions used in the equations~\eqref{eqn:pre_synaptic_ne6} to~\eqref{eqn:pre_synaptic_ne8} are
\begin{equation}
\begin{split}
    m_{\infty,a} &= \text{Hill} (p_a, d_1), \\ 
    n_{\infty,a} &= \text{Hill}(c_a,d_5), \\ 
    \text{Hill}(x^n,K) &= \frac{x^n}{x^n + K^n}, \\
    \alpha_{h_a} &= a_2 d_2 \frac{p_a+d_1}{p_a + d_3}, \\
    \beta_{h_a} &= a_2 c_a,
\end{split}
\end{equation}
where the generic Hill function $\text{Hill}(x^n, K)$ is used for reactions, whose reaction velocity curve is not hyperbolic~\citep{keener2009mathematical}.
Various studies have shown that astrocytes release gliotransmitters in a Ca$^{2+}$-dependent manner~\citep{fellin2009communication,montana2006vesicular}. The released gliotransmitters modulate synaptic plasticity through extra-synaptic NMDA receptors~\citep{parpura1994glutamate,parpura2000physiological} and extra-synaptic metabotropic Glu-receptors~\citep{perea2007astrocytes}. The release mechanism of gliotransmitters from astrocytes, similar to that of neurons, occurs in a vesicular manner~\citep{montana2006vesicular,bezzi2004astrocytes}. Studies indicate that the Ca$^{2+}$ dependence of Glu release from hippocampal astrocytes and found that the Hill coefficient for this process ranged between $2.1$ and $2.7$, indicating that at least two Ca$^{2+}$ ions are essential for gliotransmitter release (\emph{cf}.~\citep{parpura2000physiological}). Consequently, the probability of vesicular fusion in response to mechanical stimulation and the size of the readily releasable pool of synaptic-like micro-vesicles in astrocytes was determined in~\citep{malarkey2011temporal}. Based on the work of Bertram et al.~(\emph{cf}.~\citep{bertram1996single}), the model describing gliotransmitter release site activation requires that three Ca$^{2+}$ ions must bind to three independent gates or sites ($S_1-S_3$) to enable gliotransmitter release, which is expressed as
\begin{equation}
    C_a + C_j \overset{k^+_j}{\underset {k^-_j}{\rightleftharpoons}} O_j, \qquad j\in(1,2,3),
\end{equation}
with $C_j, O_j$ as the closing and equal opening probabilities of gate $S_j$, and, $k_j^+, k_j^-$, as the opening and closing rates of the gate $S_j$, respectively. The temporal evolution of the open gate $O_j$ can be expressed as
\begin{equation}
    \frac{\mathrm{d}O_j}{\mathrm{d}t} = k^+_j c_a - (k^+_j c_a + k^-_j ) O_j.
\end{equation}
The fraction of synaptic-like micro-vesicles, $f_r^a$, ready to be released is described by the product of the opening probabilities, $O_j$ with $j\in(1,2,3)$, of the three sites due to their physically independency. The dissociation constants of gates $S_1, S_2, S_3$ are $\SI{108}{\nano\M}$, $\SI{400}{\nano\M}$ and $\SI{800}{\nano\M}$. The time constants for gate closure, $1/k_j^-$, are $\SI{2.5}{\second}$, $\SI{1}{\second}$ and $\SI{100}{\milli\second}$. $S_1$ and $S_2$, as well as the time constants, are adapted from~\citep{bertram1996single}, while the dissociation and time constant for gate $S_3$ is determined iteratively to match the experimental data from~\citep{malarkey2011temporal}. When the synaptic-like micro-vesicle is ready to be released, the fusion and recycling process is modeled by the Tsodyks-Markram model~\citep{tsodyks1997neural}, given by
\begin{equation}
\begin{split}
    \frac{\mathrm{d}R_a}{\mathrm{d}t} &= \frac{I_a}{\tau^a_{rec}}- \Theta (c_a - c_a^{thresh})f_r^a R_a, \\
    \frac{\mathrm{d}E_a}{\mathrm{d}t} &= - \frac{E_a}{\tau_{inact}^a} + \Theta (c_a - c_a^{thresh})f_r^a R_a, \\
    I_a &= 1 - R_a - E_a.
\end{split}
\end{equation}
In this context, $R_a$ denotes the fraction of readily releasable synaptic-like micro-vesicles within the astrocytes, $E_a$ represents the fraction of effective synaptic-like micro-vesicles in the extra-\ac{SMCs}, and $I_a$ signifies the fraction of inactive synaptic-like micro-vesicles undergoing endocytosis or re-acidification. The term $\Theta$ is the Heaviside function, and $c_a^{thresh}$ represents the threshold of astrocytes [Ca$^{2+}$] required for activation of the release site~\citep{parpura2000physiological}. The time constants for the inactivation and recovery of synaptic-like micro-vesicles are denoted by $\tau^a_{inact}$ and $\tau^a_{rec}$, respectively.

\subsection{Components of the Neurovascular Coupling Model}
To extend the quadripartite synapse model into a NGVU model, it is essential to incorporate smooth muscle and endothelial cell contraction, as well as vascular mechanisms, to achieve completeness. For this purpose, we employ the neurovascular coupling model developed in~\citep{dormanns2015neurovascular}, with necessary modifications to address certain limitations in the original. For instance, the original model incorrectly assumes that astrocytes exhibit action potentials, which is not supported by current understanding. Additionally, it considers potassium as the primary driving component in neurons and astrocytes, an assumption that has been updated in light of recent findings~\citep{brazhe2023astrocytes}. The neurovascular unit model described in~\citep{dormanns2015neurovascular} comprises seven fundamental compartments: neurons, the synaptic cleft, astrocytes, the perivascular space, \ac{SMCs}, \ac{ECs}, and the arteriolar lumen. These compartments are interconnected through four subsystems: (1) the neuron/astrocyte subsystem, which includes the perivascular space and the synaptic cleft, (2) the smooth muscle cell/endothelial cell subsystem that couples these two cell types, (3) the arteriolar contraction subsystem, and (4) the arteriolar wall mechanical subsystem. The subsystems are modeled as having a triggering input and a subsequent output, which establishes connectivity and serves as a triggering input for other interconnected subsystems~\citep{dormanns2015neurovascular}. Given the incorrect assumptions about the neuron/astrocyte subsystem in the original model, we include only the latter three subsystems in the NGVU model.

\subsubsection{Smooth Muscle and Endothelial Cell Dynamics}\leavevmode\\
The subsystem combining the smooth muscle cell and endothelial cell is based on the work in~\citep{koenigsberger2006effects}. In the original model of~\citep{dormanns2015neurovascular}, an \ac{KIR} channel at the interface between the smooth muscle cell and the perivascular space is included. The inputs for this subsystem are the KIR channel on the smooth muscle cell facing the perivascular space, which allows a flux of K$^+$ into the cytosol, and the influx of IP$_3$ into the endothelial cell mediated by luminal agonist P2Y receptors on the endothelial cell membrane~\citep{dormanns2015neurovascular}. The biochemical behavior of the KIR channel is modeled and implemented based on experimental data from~\citep{filosa2006local} as
\begin{equation}
    J_{KIR_i} = \frac{F_{KIR_i} g_{KIR_i}}{\gamma_i} (v_i - v_{KIR_i}),
\end{equation}
where the conversion parameter $\gamma_i$ is used to provide $J_{KIR_i}$ in the correct units of $\SI{}{\milli\volt\per\second}$. The Nernst potential (in $\SI{}{\milli\volt}$) is a function of $K_p$ and is expressed as 
\begin{equation}
    v_{KIR_i} = z_1 K_p - z_2,
\end{equation}
where $z_1$ and $z_2$ are derived by fitting a linear function according to~\citep{filosa2006local}. The conductance of the KIR channel depends on both the membrane potential $v_i$ and $K_p$. The values of the constants are likewise based on the experimental data from~\citep{filosa2006local}. The second input to the smooth muscle cell/endothelial cell subsystem is the generation of IP$_3$ in the endothelial cell due to the activation of membrane receptors by agonists in the arteriolar lumen. IP$_3$ mediates the $J_{IP_3}$ channel in both the endothelial cell and smooth muscle cell, located on the surface of the endothelial cell and saroplasmatic reticulum, allowing Ca$^{2+}$ to be released from the reticulum. For a given IP$_3$ concentration in the smooth muscle cell/endothelial cell subsystem, Ca$^{2+}$ fluctuations can occur due to the CICR process~\citep{goldbeter1990minimal, koenigsberger2005role}. The production of IP$_3$ in endothelial cell is represented by the flux $J_{EC,IP_3}$. In this study, IP$_3$ production is constant but can be extended to a time-dependent function.
\\
\indent Physiologically, endothelial and \ac{SMCs} are connected by hetero- and homocellular gap junctions that allow intercellular exchange of molecules and voltage. In the neurovascular unit model in~\citep{dormanns2015neurovascular}, the endothelial and \ac{SMCs} are modeled as a single subsystem, thus the homocellular exchange between cells is neglected. The linearized coupling functions for the heterocellular exchange of calcium, voltage, and IP$_3$ are given by
\begin{align}
    J_{Ca^{2+}_{cpl}} &= -P_{Ca^{2+}} ([Ca^{2+}]_i - [Ca^{2+}]_j), \\
    J_{V_{cpl}} &= -G_v (V_i - V_j), \\
    J_{IP_{3,{cpl}}} &= -P_{IP_3} ([IP_3]_i-[IP_3]_j),
\end{align}
where the subscripts $i$ and $j$ correspond to endothelial and \ac{SMCs}, respectively. Additionally, calcium buffering is incorporated to account for the fact that approximately 1\% of the intracellular Ca$^{2+}$ is free~\citep{parthimos1999minimal}.

\subsubsection{Arteriolar Contraction Model}\leavevmode\\
The contraction force is generated through the formation of cross-bridges between actin and myosin filaments, a process regulated by cytosolic Ca$^{2+}$. The model for the arteriolar contraction subsystem builds upon the foundational work of Hai and Murphy~\citep{hai1989ca2+}. The input signal for this model is the compartmental cytosolic Ca$^{2+}$ concentration in \ac{SMCs}. Myosin exists in four potential states during this process: free non-phosphorylated cross-bridges (M), free phosphorylated cross-bridges (Mp), attached phosphorylated cross-bridges (AMp), and attached dephosphorylated latch-bridges (AM). The dynamics of the fraction of myosin in each state are described by the following four equations as
\begin{align}
    \frac{\mathrm{d}[Mp]}{\mathrm{d}t} &= K_4 [AMp] + K_1 [M] - (K_2 + K_3) [Mp], \\
    \frac{\mathrm{d}[AMp]}{\mathrm{d}t} &= K_3 [Mp] + K_6 [AM] - (K_4 + K_5) [AMp], \\
    \frac{\mathrm{d}[AM]}{\mathrm{d}t} &= K_5 [AMp] - (K_7 + K_6) [AM], \\
    [M] &= 1 - [AM] - [AMp] - [Mp],
\end{align}
where the rate constants $K_n$ with $n=(1,...,7)$ control the phosphorylation and bridge formation processes. The Ca$^{2+}$ dependence of the cross-bridge model is governed by the rate constants $K_1$ and $K_6$. According to~\citep{koenigsberger2006effects}, the total phosphorylation of myosin is a function of the smooth muscle cell compartmental Ca$^{2+}$. Thus, $K_1$ and $K_6$ are defined by the constants characterizing the Ca$^{2+}$ sensitivity of calcium-activated phosphorylation of myosin as
\begin{equation}
    K_1 = K_6 = \gamma_{cross} [Ca^{2+}]^3_i.
\end{equation}
The active stress in a smooth muscle cell is directly proportional to the fraction of attached cross-bridges, $F_r$, which serves as an input parameter for the mechanical subsystem. The fraction of attached cross-bridges is mathematically described as follows
\begin{equation}
\label{eq:fr}
    F_r = \frac{[AMp]+[AM]}{\left([AMp]+[AM]\right)_{max}}.
\end{equation}
$\left([AMp] + [AM]\right)_{max}$ represents the maximum possible value of the sum of the concentrations of attached phosphorylated ($[AMp]$) and dephosphorylated ($[AM]$) cross-bridges. It is used for normalization, ensuring that $F_r$ is dimensionless and scaled between 0 and 1, which facilitates its use as an input parameter for the mechanical subsystem.

\subsubsection{Visco-Elastic Model of the Arteriolar Wall}\label{awms}\leavevmode\\
The mechanical properties of the arterial wall are assumed as visco-elastic and are modeled following the Kelvin-Voigt approach. This model comprises a Newtonian damper and a Hookean elastic spring connected in parallel. The input signal, which corresponds to the active stress state of the smooth muscle cell in the circumferential direction, is the fraction of attached myosin cross-bridges, $F_r$, from the mechanical subsystem. The circumferential stress in the arterial wall, $\sigma_{\theta \theta}$, is given by the linear elastic part with the Young's modulus $E$, and the viscous part with the viscosity $\eta$, as
\begin{equation}
    \sigma_{\theta \theta} = E \epsilon_{\theta \theta} + \eta \frac{\mathrm{d}\epsilon_{\theta \theta}}{\mathrm{d}t}=\frac{R\Delta p}{h},
\end{equation}
where  $\epsilon_{\theta \theta}$ is the strain in the arterial wall, $\Delta p$ represents the transmural pressure, $R$ denotes the vessel radius, and $h$ is the vessel thickness. For simplicity, the wall thickness in this simulation is considered a constant fraction of the radius, with $h=0.1R$. The parameters of this subsystem are adapted from experimental data available in~\citep{davis1985capillary}. A linear function is employed to map the transition from the fully activated state to the fully relaxed state, applicable to radii between $\SI{10}{\micro\metre}$ and $\SI{30}{\micro\metre}$. The Young's modulus as well as the initial radius are assumed to be continuous functions of $F_r$, with a linear interpolation between the active ($E_{act}$) and passive ($E_{pas}$) experimental data points provided in~\citep{davis1985capillary}, expressed as
\begin{equation}
    E(F_r) = E_{pas} + F_r (E_{act}-E_{pas}).
\end{equation}
This approach can similarly be applied to represent the initial radius, yielding
\begin{equation}
    R_0 (F_r) = R_{0,pas}+F_r(R_{0,act}-R_{0,pas}).
\end{equation}
Transforming the latter equations into a time-dependent expression for the vessel radius results in
\begin{equation}
    \label{eqn:dR_dt}
    \frac{\mathrm{d}R}{\mathrm{d}t} = \frac{R_{0,pas}}{\eta} \left(\frac{R\Delta p}{h}-E(F_r) \frac{R-R_0(F_r)}{R_0(F_r)}\right).
\end{equation}
The final equation~\eqref{eqn:dR_dt} describes the dynamic behavior of the vessel radius as a function of the transmural pressure, visco-elastic properties of the wall, and the activation state of \ac{SMCs} via $F_r$. This expression expressed the connection between mechanical forces and active stress, allowing for a time-dependent simulation of arteriolar behavior under varying physiological conditions, providing a model for the dynamic regulation of blood flow and vessel radii in response to changes in transmural pressure and smooth muscle activation.

\subsubsection{Influence of the Nitric Oxide Signaling Pathway}\leavevmode\\
\ac{NO} plays a fundamental role in neurovascular coupling by mediating the dynamic regulation of vascular tone in response to neuronal activity. As a potent vasodilator, \ac{NO} is critical for linking cellular signaling processes to changes in arteriolar diameter and cerebral blood flow. Its high diffusivity and involvement in complex biochemical pathways necessitate its inclusion in the neurovascular coupling model to achieve a mechanistic understanding of the connection between neuronal activation, vascular responses, and metabolic demands~\citep{dormanns2016role}. The \ac{NO} signaling pathway model is based on~\citep{dormanns2016role}, which is an addition to their previously introduced neurovascular unit model~(\emph{cf}.~\citep{dormanns2015neurovascular}). The \textbf{NO} signaling pathway is mathematically described by production, diffusion and consumption terms in different cells, as well as the interaction of \ac{NO} with other biochemical species and ion channel open probabilities~\citep{dormanns2016role}.
\\
\indent The production of \textbf{NO} is driven by the two constitutive isoforms of \ac{NOS}, \ac{nNOS} and \ac{eNOS}~\citep{fleming1999NO}. The activation of both enzymes is mediated by intracellular Ca$^{2+}$ in the neurons and \ac{ECs}, respectively. Additionally, \ac{eNOS} is activated by blood flow-induced wall shear stress in cerebral arterioles~\citep{joannides1995nitric}. Due to its high diffusion coefficient, \ac{NO} rapidly diffuses into other compartments, as demonstrated by experimental data~\citep{malinski1993nitric} and kinetic simulations~\citep{lancaster1994simulation}. When \ac{NO} reaches the smooth muscle cell, it interacts with intracellular enzymes to regulate smooth muscle cell relaxation~\citep{yang2005mathematical}. The dynamics of \ac{NO} in the various compartments are described using mass balance equations. The concentration of \ac{NO} in each domain is determined by the balance between production, intracellular consumption, and diffusion to and from other compartments. Fig.~\ref{fig:NGVUmodel} graphically illustrates the \ac{NO} signaling pathway.

The rate of NO production is determined by the concentration of activated NO synthases. The essential biochemical substrates for this process include \ac{L-Arg}, \ac{O$_2$}, and \ac{NADPH}~\citep{chen2006theoretical}, with \ac{L-Arg} serving as the crucial nitrogen donor~\citep{forsythe2001mast}. During the reaction, L-Arg is converted into L-citrulline, producing \ac{NADP}$^+$, water, and NO in the process. This reaction involves a five-electron oxidation that occurs in two distinct steps, and the overall stoichiometric equation is given by
\begin{equation}
    \text{L-Arg} + 2\text{O}_2 + \frac{2}{3} \text{NADPH} \overset{NOS}{\longrightarrow} \text{L-Citrulline} + \frac{2}{3} \text{NADP}^+ + 2\text{H}_2\text{O} + \text{NO}.
\end{equation}
Several biomolecule cofactors, including flavin mononucleotide, flavin adenine dinucleotide, and tetrahydrobiopterin, are required for the reaction. The constitutive NOS isoforms, nNOS and eNOS, are considered the primary producers of NO and play a crucial role in maintaining homeostasis~\citep{fleming1999NO}. Consequently, in this model, NO production is limited to neurons and \ac{ECs}, with the production rates in astrocytes ($p_{NO,k}$) and \ac{SMCs} ($p_{NO,i}$) set to zero. In neurons, the synthesis of NO is catalyzed by nNOS, which is activated in response to glutamate-induced calcium influx into the post-synaptic neurons. The production of NO in neurons depends on the concentration of activated nNOS and is further constrained by the availability of O$2$ and L-Arg~\citep{chen2006theoretical}. This relationship is expressed mathematically through the maximum neuronal production rate, $p{max}$, as
\begin{align}
\begin{split}
    p_{NO,n} &= p_{max} \frac{[O_2]_n}{K_{m,O2,n}+[O_2]_n} \frac{[L-Arg]_n}{K_{m,L-Arg,n}+[L-Arg]_n}, \\
    p_{max} &= V_{max,NO,n} [nNOS_act]_n.
\end{split}
\end{align}
The maximum catalytic rate of nNOS, denoted as $V_{max,NO,n}$, along with the neuronal concentrations of O$_2$ and L-Arg, $[O_2]_n$ and $[L-Arg]_n$, respectively, and the Michaelis constants $K_{m,O2,n}$ and $K_{m,L-Arg,n}$, define the reaction kinetics (\emph{cf}.~\citep{chen2006theoretical}). The activation of nNOS in the neurons is induced by the neurotransmitter Glu in response to neuronal activation. In a chemical synapse, when an action potential arrives at the presynaptic neuronal axon terminal, Glu is released from vesicles into the synaptic cleft. This Glu then binds to receptors on the dendrite of the postsynaptic neurons and is subsequently cleared from the synaptic cleft through diffusion and hydrolysis~(\emph{cf}.~\citep{dormanns2016role,brazhe2023astrocytes,tewari2012mathematical}), as described in the quadripartite synapse model in Section~\ref{sec:glu_gaba}.

Since the neuronal NOS is often co-localized with ionotropic NMDA-receptors~\citep{mayer1997biosynthesis, forstermann1998expressional}, which are receptor complexes including transmembrane ion channels in the neurons that are opened or closed in response to the binding of Glu~\citep{benarroch2006basic}. There are two different types of NMDA-receptors, NR2A and NR2B, which show different opening probabilities, $w$, which depend on the neuronal Glu concentration using Michaelis-Menten kinetics to meet the experimental data from~\citep{santucci2008effects}. Mathematically, the opening probability $w$ is described by 
\begin{equation}
    w_{NR2,i} = \frac{[Glu]_{sc}}{K_{m,i}+[Glu]_{sc}}, \qquad i \in \{A,B\},
\end{equation}
where $K_{m,A}$ and $K_{m,B}$ are the fitted Michaelis constants. Given that NMDA receptors are highly permeable to Ca$^{2+}$ ions (\emph{cf}.~\citep{jahr1993calcium}), the glutamate-induced Ca$^{2+}$ flux is incorporated into the model. The total calcium current, $I_{Ca,tot}$, is expressed as 
\begin{align}
\begin{split}
    I_{Ca,tot} &= I_{Ca} (0.63 w_{NR2,A}+11 w_{NR2,B}) \\
    I_{Ca} &= \frac{4v_nG_M(P_{Ca}/P_M) ([Ca^{2+}]_{ex}/[M])}{1+\exp(\alpha_v(v_n+\beta_n))} \frac{\exp(2v_n F/(R_{gas}T))}{1-\exp(2v_nF/(R_{gas}T))},
\end{split}
\end{align}
where $I_{Ca}$ is the neuronal inward calcium current, $F$ is Faraday's constant, $v_n$ is the neuronal membrane potential, $G_M$ is the conductance, and $P_{Ca}/P_M$ is the ratio of NMDA receptor permeabilities to Ca$^{2+}$ and monovalent ions. Additionally, the model accounts for the external calcium concentration $[Ca^{2+}]_{ex}$, the concentration of monovalent ions $[M]$, and the intra- and extracellular translation factors $\alpha_v$ and $\beta_v$. The temperature $T$ and the universal gas constant $R_{gas}$ are also included. The estimated arrival of 0.63 NR2A- and 11 NR2B-NMDA-receptors, on average per synapse, is based on measurements from~\citep{santucci2008effects}. In the endothelial cell, NO production is facilitated by the constitutive enzyme isoform eNOS, whose catalytic activity depends on the availability of O$_2$ and L-Arg. The endothelial NO production can be described by
\begin{equation}
    p_{NO,j} = V_{max,NO,j} [eNOS_{act}]_j \frac{[O_2]_j}{K_{m,O2,j} + [O_2]_j} \frac{[L-Arg]_j}{K_{m,L-Arg,j}+[L-Arg]_j}.
\end{equation}
The maximal activity of eNOS is regulated by the intracellular calcium concentration and is further influenced by wall shear stress, which arises from blood flow through the perfusing arteriole. Wall shear stress activates a signaling cascade involving phosphatidylinositol 3-kinase and the serine/threonine-specific protein kinase, which phosphorylates eNOS~\citep{chen2006theoretical, comerford2008endothelial, dormanns2016role}. The elastic strain energy stored within the vessel membrane, as well as the quantification of exogenous Ca$^{2+}$ entry via shear stress-gated ion channels, is detailed in~\citep{comerford2008endothelial}. The strain energy stored in the membrane, $F_{wss}$, and the strain energy density, $W_{wss}$, are described mathematically in~\citep{wiesner1997mathematical}. Wall shear stress, $\tau_{WSS}$, represents the frictional force per unit area and establishes a positive feedback mechanism: increased wall shear stress enhances NO production by \ac{ECs}, inducing vasodilation and further modifying the shear stress. The wall shear stress $\tau_{WSS}$ in the arteriolar wall depends on the regional \ac{CBF}, $Q$, which is modeled using the Hagen–Poiseuille flow assumption for cerebral arterioles.

The diffusion of NO, denoted as $d_{NO,m}$, can be derived from Fick's second law of diffusion, which characterizes the diffusion of a substance over time and space. To simplify the diffusion formulation, the Einstein-Smoluchowski equation is employed, describing the characteristic time required for NO to diffuse a certain distance from the center of one cell to another. The compartment model outlines the diffusion between different domains but does not account for the amount of NO released into the lumen and subsequently scavenged by reactions with hemoglobin. The diffusion term of a linear diffusion is given by
\begin{align}
    d_{NO,m} &= \frac{[NO]_{out} - [NO]_{in}}{\tau_{\Delta x, m}} , \qquad m \in \{n,a,s,e\}, \\
    \tau_{\Delta x,m} &= \frac{(\Delta x_m)^2}{2D_{c,NO}},
\end{align}
where $m \in \{n,a,s,e\}$ denotes the cell indices for neurons, astrocytes, smooth muscle and \ac{ECs}, respectively, $\tau_{\Delta_x}$ represents the characteristic time required for NO to diffuse over a specific distance $\Delta x_m$ from the center of one cell to another. For example, the distances between the centers of the neurons and astrocytes is roughly $\SI{25}{\micro\metre}$~\citep{dormanns2016role, kavdia2002model}.

The consumption of NO is denoted as $c_{NO,m}$. As a free radical, NO readily reacts with biochemical species containing unpaired electrons, such as molecular O$_2$, superoxide anions, and metals~\citep{mayer1997biosynthesis}. NO is scavenged in the cytosol of all cell types through which it diffuses. The mathematical expression for the scavenging term across all model compartments is given by:
\begin{equation}
c_{NO,m} = k_m [NO]_m C_m , \qquad m \in \{n,a,e\},
\end{equation}
where $C_m$ is the concentration of reactive species in the cell type, and $k_m$ is the reaction rate constant~\citep{kavdia2002model}. In the smooth muscle cell, NO affects the contraction mechanism and the open probability of the large conductance Ca$^{2+}$-dependent K$^+$ channel (BK) channel via its second messenger, \ac{cGMP}. Additionally, NO activates soluble guanylyl cyclase, an enzyme that catalyzes the formation of \ac{cGMP}. The kinetics of these processes are described in~\citep{yang2005mathematical}.

\subsection{Integration of Synapse and Vascular Components in NGVU}
\label{CombNGVU}
\begin{figure}[htbp]
    \centering
        \includegraphics[width=\linewidth]{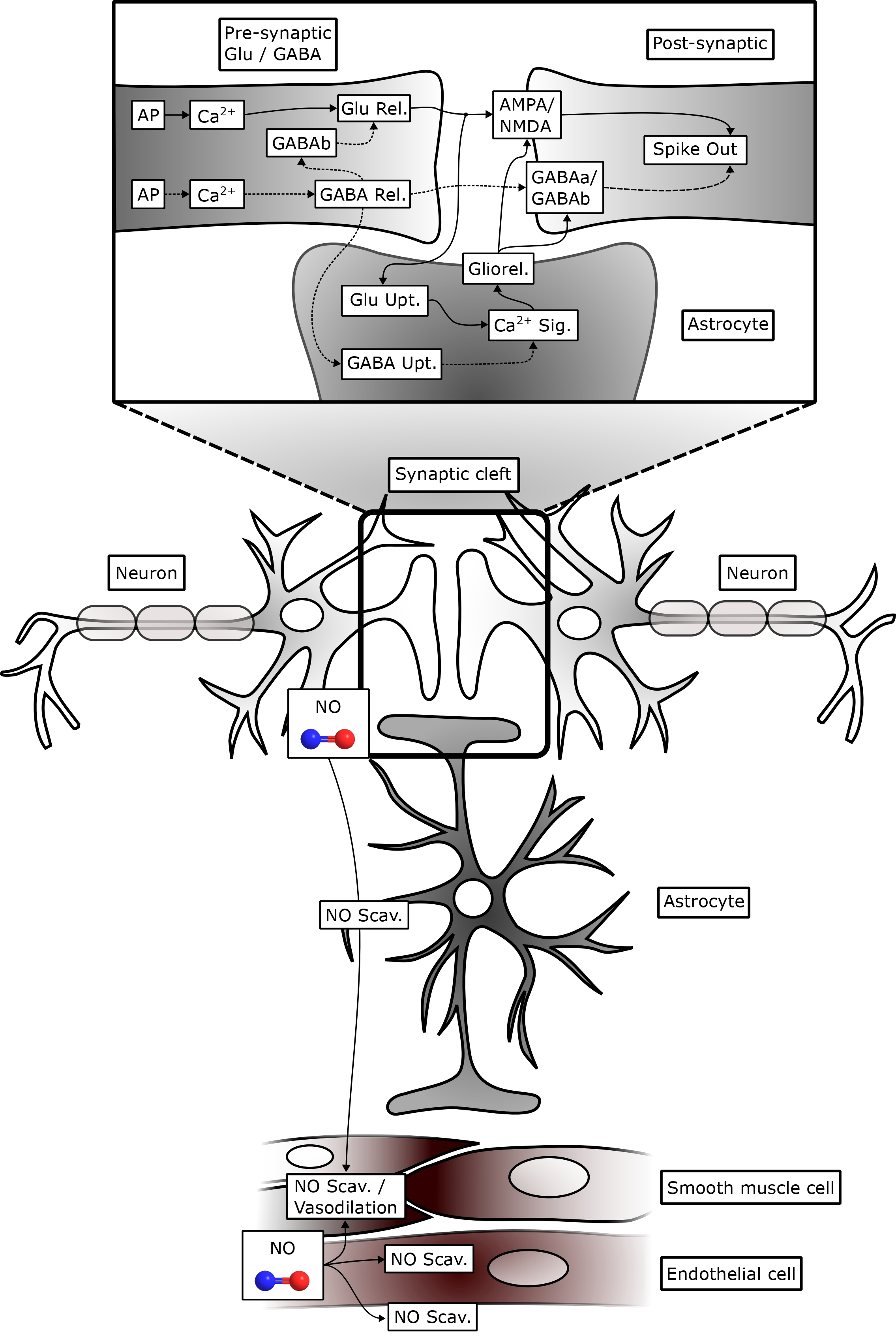}
    \caption{    Components of the NGVU. Quadripartite synapse with pre-synaptic Glu/GABA neurons (top left), post-synaptic neuron (top right), 
    astrocyte (center). Action potentials (AP) lead to Ca\textsuperscript{2+} influx, triggering 
    Glu Rel/GABA Rel into the synaptic cleft. The astrocyte uptakes neurotransmitters (Glu Upt., GABA Upt.) 
    and, via Ca\textsuperscript{2+} Sig., releases gliotransmitters (Gliorel). Endothelial and \ac{SMCs} 
    (bottom) regulate vasodilation through NO production and NO scavenging (Scav.). Abbreviations: AMPA/NMDA – excitatory receptors, 
    GABA\textsubscript{a}/GABA\textsubscript{b} – inhibitory receptors, Spike Out – post-synaptic spiking.    }
    \label{fig:NGVUmodel}
\end{figure}

To integrate the quadripartite synapse and vascularization models into a coherent NGVU model, the interaction points between subsystems must be carefully addressed. First, the neurons/astrocytes subsystem from the original neurovascular unit model in~\citep{dormanns2015neurovascular} is extended with the quadripartite synapse model, while the vascularization components retain the smooth muscle cell/endothelial cell model, as well as the arteriolar contraction and mechanical wall models. Specifically, the astrocytic Ca$^{2+}$ released over time is incorporated into the sarcoplasmic reticulum as an uptake process. The Ca$^{2+}$ flux into the sarcoplasmic reticulum is modeled a 
\begin{equation}
    J_{SR_{uptake,i}}=B_i \frac{[Ca^{2+}]^2_i}{c^2_{bi}+[Ca^{2+}]^2_i},
\end{equation}
where $[Ca^{2+}]_i$ represents the time-dependent Ca$^{2+}$ dynamics in astrocytes as defined by the quadripartite synapse model. The Ca$^{2+}$ release through the sarcoplasmic reticulum leak channel is driven by IP$_3$ generation. The initial input to the smooth muscle cell/endothelial cell subsystem through the KIR channel is retained in the NGVU model. Fig.~\ref{fig:NGVUmodel} provides an overview of the complete NGVU model, integrating the quadripartite synapse and vascularization components, along with the relevant channels and fluxes.

To integrate the gasotransmitter NO into the NGVU model, Yang et al.~\citep{yang2005mathematical} proposed two pathways through which NO can induce local vasodilation of arterioles in the brain. The first pathway involves an indirect influence on the smooth muscle cell contractile system, which is governed by the formation of cross-bridges between actin and myosin filaments, as initially described by Hai and Murphy~\citep{hai1988cross}. Furthermore, cGMP exerts a regulatory effect on the myosin dephosphorylation process, altering the rate constants for dephosphorylation as described by Yang et al.~\citep{yang2005mathematical}, represented by
\begin{align}
    R_{cGMP}&=\frac{[cGMP]_i^2}{K^2_{m,mlcp}+[cGMP]^2_i}, \\
    K_{2c} &= K_{5c} = \delta_i (k_{mlpc,b}+k_{mlpc,c}R_{cGMP}),
\end{align}
where $K_{m,mlcp}$ denotes the Michaelis constant, $k_{mlpc,b}$ is the basal myosin light-chain dephosphorylation rate constant, $k_{mlpc,c}$ is the first-order rate constant for cGMP-regulated myosin light-chain dephosphorylation, and $\delta_i$ is a constant fitted to experimental data (\emph{cf}.~\citep{yang2005mathematical, hai1988cross}). The second messenger of NO, cGMP, modifies the rate constants for the dephosphorylation of $Mp$ to $M$ and $AMp$ to $AM$ via myosin light-chain phosphatase~\citep{yang2005mathematical}. The second pathway concerns the open probability, $w_i$, of the BK channel in \ac{SMCs}, which is a function of the membrane potential, $v_i$, and is shifted to the left in the membrane potential space by cGMP (\emph{cf}.~\citep{stockand1996mechanism}). Instead of using the open probability of the BK channel as described by Koenigsberger et al.~\citep{koenigsberger2006effects}, the influence of NO is incorporated as a cGMP-dependent modification of the open probability, $w_i$. Mathematically, this is expressed as
\begin{equation}
    c_{w,i} = \frac{1}{\epsilon_i + \alpha_i \exp(\gamma_i[cGMP]_i)},
\end{equation}
where $\epsilon_i$, $\alpha_i$, and $\gamma_i$ are translation factors calibrated to match the observed data from Stockand and Sansom~\citep{stockand1996mechanism}. Furthermore, to fully integrate the NO model into the NGVU model, we use the Glu in the synaptic cleft, provided by the quadripartite synapse model.

\section{Numerical Simulation Results}\label{sec:results}
\begin{figure}[htbp!]
  \centering
  \includegraphics[width=\linewidth]{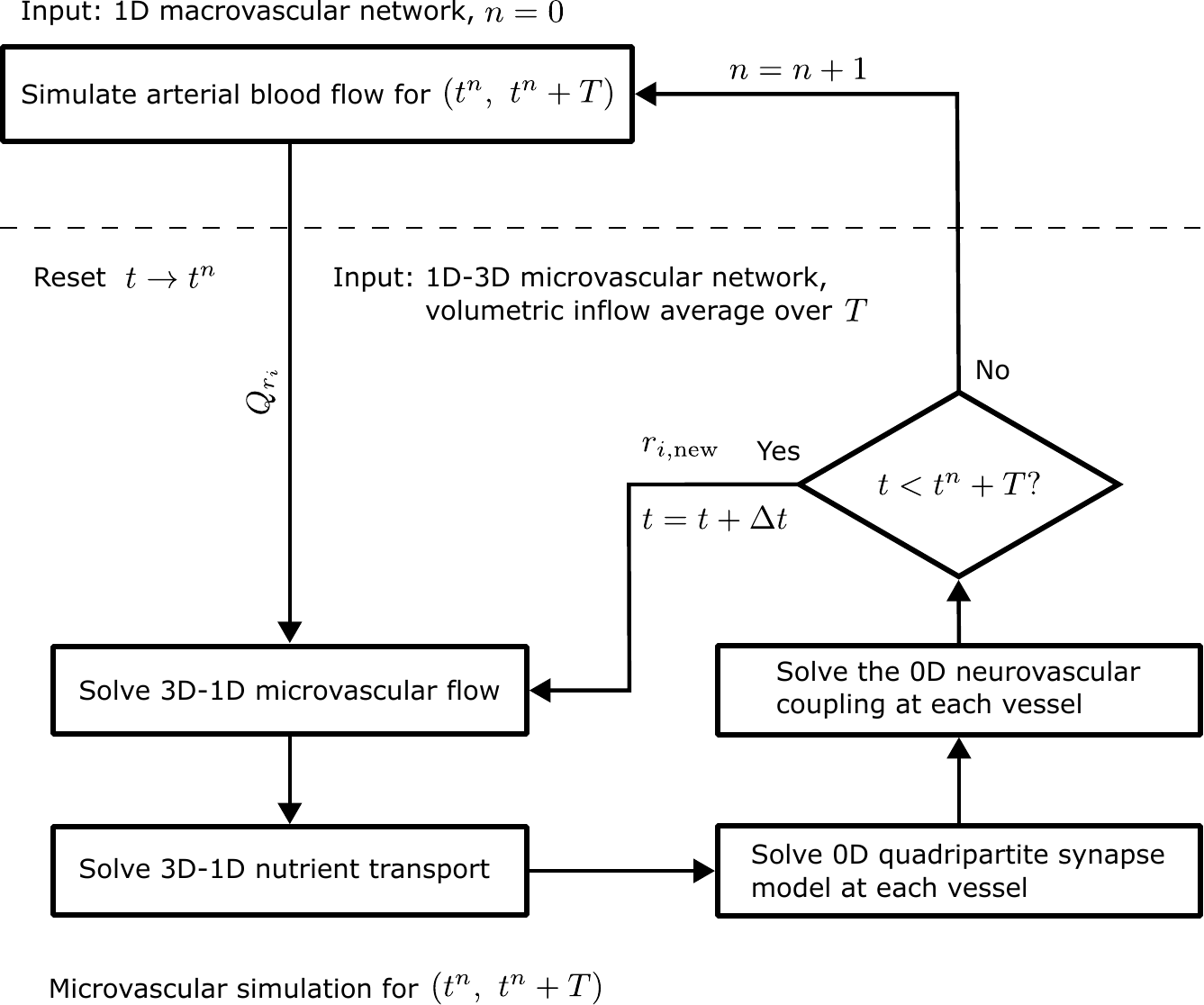}
  \caption{Schematic of the simulation workflow during one heartbeat period ($T$), illustrating the sequential coupling between the 1D macrovascular network and the neurovascular-coupled 3D-1D-0D microvascular network.}
  \label{fig:algorithm}
\end{figure}

The following sections present the computational results of our integrated neuro-glial-vascular coupling model. The findings are organized to provide a detailed examination of the macrocirculation dynamics, neurotransmitter release, neuronal spiking activity, astrocytic signaling, neurovascular responses, and microcirculation within a representative unit of the \ac{DVC}. Each subsection highlights specific aspects of the modeled system, emphasizing the interactions between neuronal, astrocytic, and vascular components and their implications for blood flow regulation and metabolic support in the \ac{DVC}. These results validate the framework and provide mechanistic insights into the physiological processes underlying neurovascular coupling. In all subsequent cellular-scale (0D) analyses, the displayed results are exemplary for the capillary vessel that receives the prescribed arterial inflow. As shown schematically in Figure~\ref{fig:algorithm}, the simulation workflow couples the macrocirculation and neurovascular microcirculation over a single heartbeat period ($T$).

\subsection{Macrocirculation Dynamics}
\begin{figure}[htb]
    \centering
    \includegraphics[width=1.0\linewidth]{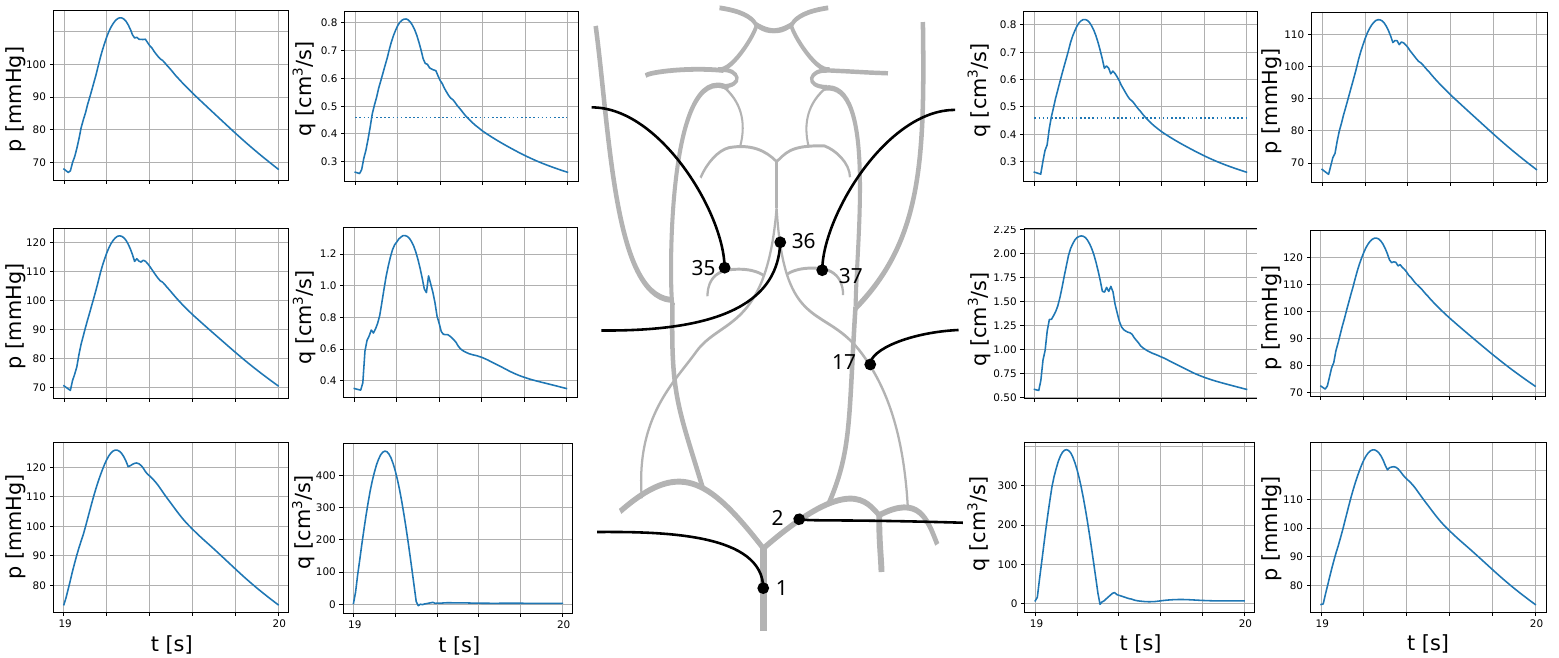}
    \caption{Flow and pressure in our Circle-of-Willis close to the coupling point with our microcirculation model.}
    \label{fig:macrocircFlowAndPressure}
\end{figure}

Figure~\ref{fig:macrocircFlowAndPressure} depicts the flow and pressure in our 1D-0D coupled model described in Section~\ref{sec:macrocirculation} at some sample points on the path from the aorta to the vessels that are coupled with the 3D-1D coupled microcirculation model (see Section~\ref{sec:microcirculation}). The average flows $Q_0$ used in the coupling are indicated by dotted lines. For the given setting assuming $\gamma = 3$ we obtain: 
$$
Q_0 \approx 0.46 \;\frac{\unit{cm^3}}{\unit{s}}.
$$
Based on this value we have for $\Lambda_c$ the following inflow rates:
$$
Q_{r_{c,1}} \approx 8.19 \times 10^{-8} \frac{\unit{cm^3}}{\unit{s}} \text{ and } Q_{r_{c,2}} \approx 5.75 \times 10^{-8} \frac{\unit{cm^3}}{\unit{s}} .
$$
The pressure curves in our macrovascular network exhibit the typical shape of pressure curves in larger arteries. It can be observed that the pressures are within a resonable range of about $70\;\unit{mmHg}$ to about $125\;\unit{mmHg}$. During the systole we have a strong increase and during the diastole a slow decay is seen. At the interface of both phases there is a characteristic kink caused by the closing of the aortic valve. Due to the Windkessel effect caused by the models attached to the outlets of the macrovascular network the flow rates do not fall back to zero apart from the aorta. This ensures a constant blood supply of the organs linked to the different outlets.
\subsection{Neurotransmitter Release and Synaptic Activation}

\begin{figure}[h!]
    \centering
    \makebox[\textwidth]{        \adjustbox{max width=\textwidth+3cm}{
            \includegraphics[width=1.2\linewidth]{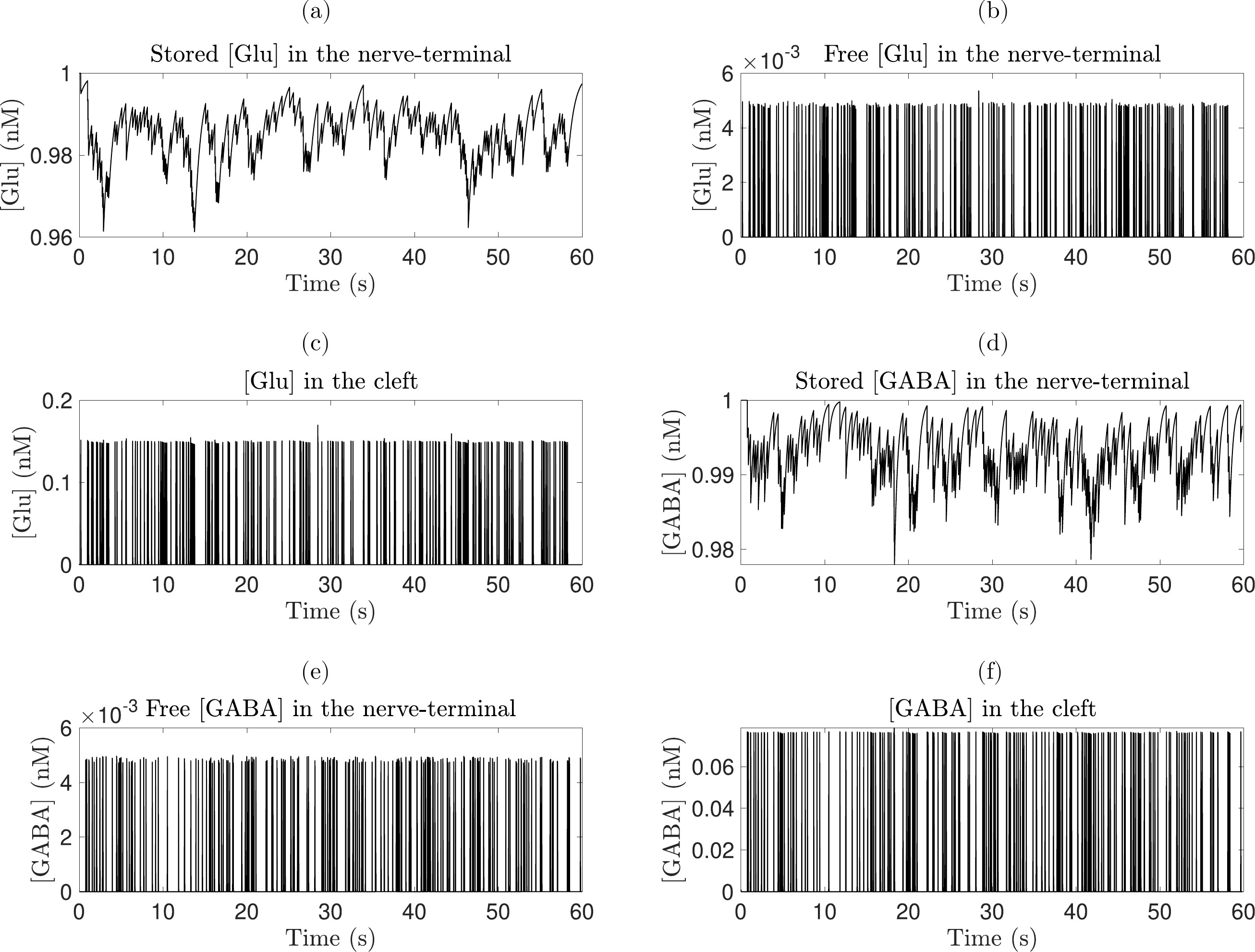}
        }
    }
    \caption{Dynamics of neurotransmitter release in a representative 0D quadripartite synapse system associated with one vessel. (a) Stored glutamate [Glu] in the nerve terminal, (b) Free glutamate [Glu] in the nerve terminal, (c) Glutamate [Glu] in the synaptic cleft, (d) Stored GABA in the nerve terminal, (e) Free GABA in the nerve terminal, and (f) GABA in the synaptic cleft.}
    \label{fig:quadripartite_synapse_results}
\end{figure}

Figure~\ref{fig:quadripartite_synapse_results} illustrates the dynamics of neurotransmitter release and synaptic activation in a representative 0D quadripartite synapse system associated with a single blood vessel. These results highlight the temporal variations in the concentrations of \ac{Glu} and \ac{GABA} within the nerve terminals and synaptic cleft, which serve as critical inputs to the neuro-glial-vascular coupling model.

In the case of glutamate, stored Glu in the nerve terminal (Figure~\ref{fig:quadripartite_synapse_results}a) gradually decreases due to synaptic activation, leading to the release of free Glu within the terminal (Figure~\ref{fig:quadripartite_synapse_results}b). A corresponding increase in glutamate concentration in the synaptic cleft is observed (Figure~\ref{fig:quadripartite_synapse_results}c), representing effective neurotransmitter signaling. Similarly, stored GABA (Figure~\ref{fig:quadripartite_synapse_results}d) transitions into free GABA (Figure~\ref{fig:quadripartite_synapse_results}e) and is subsequently released into the synaptic cleft (Figure~\ref{fig:quadripartite_synapse_results}f). These dynamics reflect the interplay of excitatory (Glu) and inhibitory (GABA) neurotransmitter systems, which are fundamental to maintaining synaptic balance and initiating downstream cellular responses.

\begin{figure}[h!]
    \centering
    \makebox[\textwidth]{        \adjustbox{max width=\textwidth+3cm}{
            \includegraphics[width=1.2\linewidth]{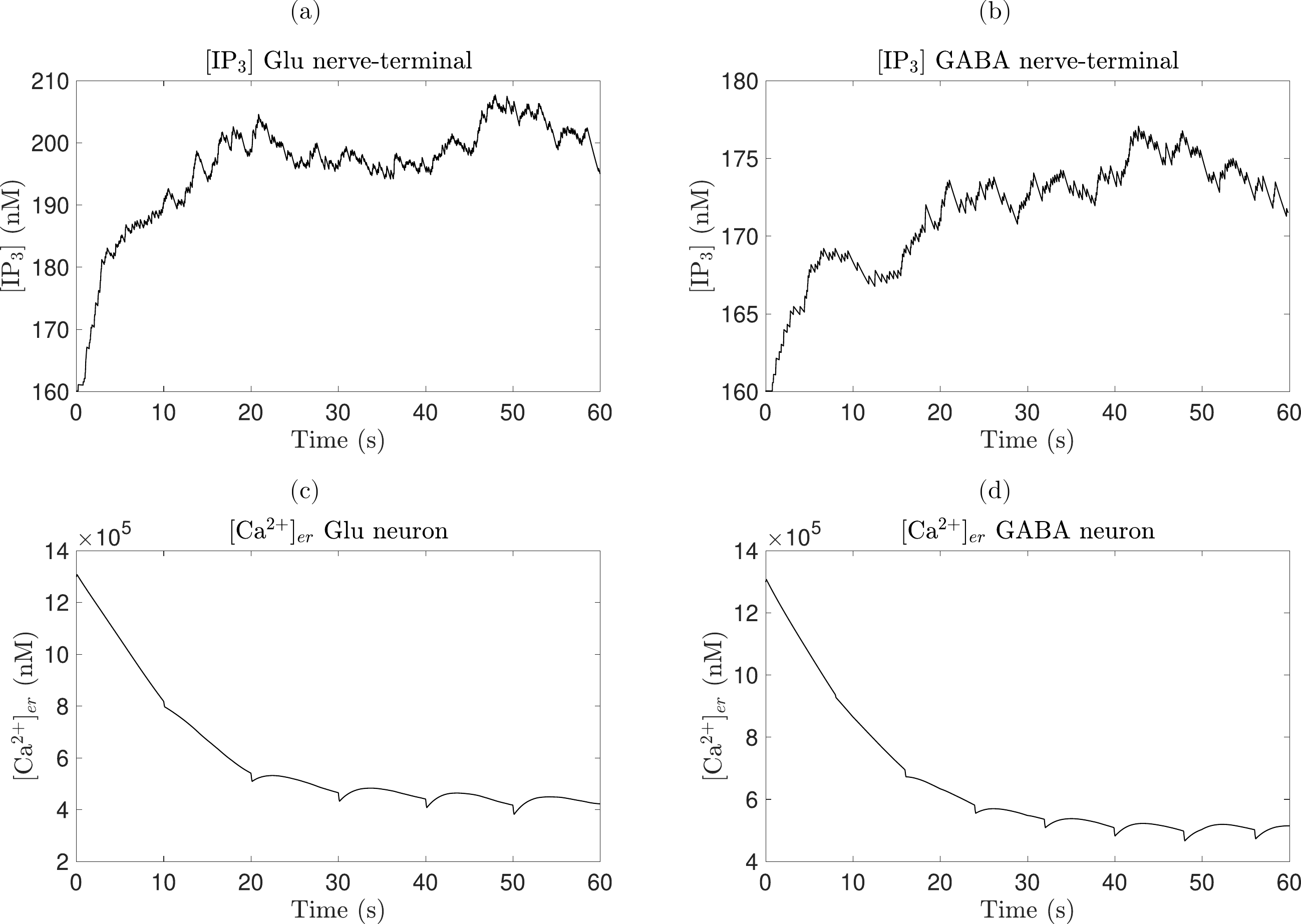}
        }
    }
    \caption{Dynamics of signaling molecules in the neuro-glial system for a representative quadripartite synapse system. (a) [IP$_3$] in the glutamate nerve terminal, (b) [IP$_3$] in the GABA nerve terminal, (c) [Ca$^{2+}$] in the \ac{ER} of the glutamatergic neuron, and (d) [Ca$^{2+}$] in the ER of the GABA neuron.}
    \label{fig:ip3_ca_results}
\end{figure}

In addition to the neurotransmitter dynamics, Figure~\ref{fig:ip3_ca_results} illustrates the temporal evolution of key signaling molecules, including inositol trisphosphate ([IP$_3$]) and calcium ion ([Ca$^{2+}$]) concentrations. The [IP$_3$] dynamics in the glutamate nerve terminal, shown in Figure~\ref{fig:ip3_ca_results}a, and the GABA nerve terminal, shown in Figure~\ref{fig:ip3_ca_results}b, play a critical role in regulating neurotransmitter release and ensuring proper synaptic function. Furthermore, the dynamics of intracellular calcium stores, depicted in Figure~\ref{fig:ip3_ca_results}c and Figure~\ref{fig:ip3_ca_results}d for the \ac{ER} of glutamate and GABA neurons, respectively, are pivotal for neurotransmitter release and downstream signaling events. These results highlight the intricate cascade of signaling events triggered by synaptic activation, emphasizing the importance of both [IP$_3$] and [Ca$^{2+}$] in maintaining synaptic activity and influencing neurovascular interactions.
Together, these figures emphasize the intricate signaling events within the quadripartite synapse model, which captures the interactions between neurons, astrocytes, and the vascular system. This representative system highlights the local coupling between synaptic activation and vascular responses, contributing to the neuro-glial-vascular coupling framework.

\subsection{Neuronal Spiking Activity}

\begin{figure}[h!]
    \centering
    \makebox[\textwidth]{        \adjustbox{max width=\textwidth+3cm}{
            \includegraphics[width=1.2\textwidth]{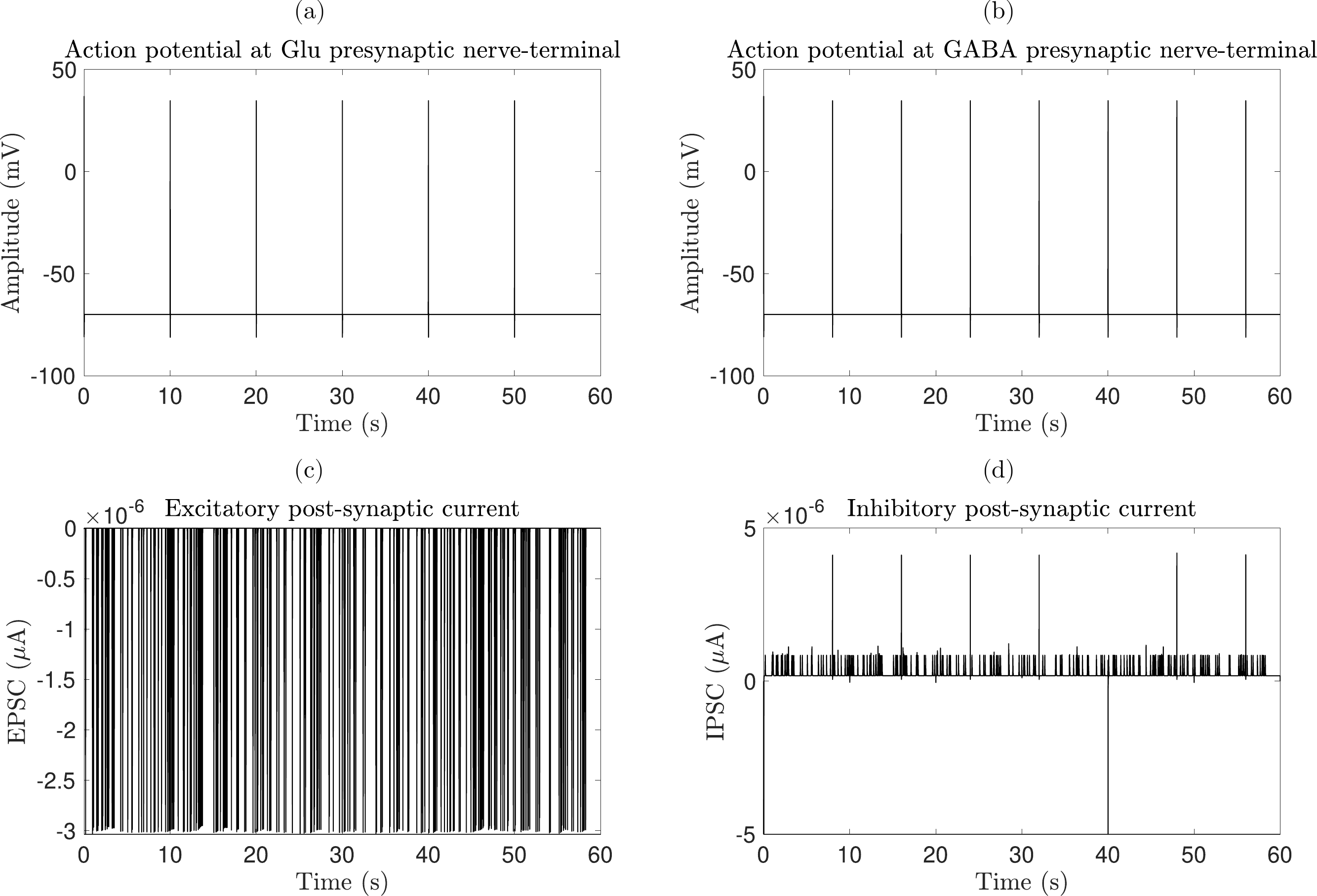}
        }
    }
    \caption{Neuronal spiking activity and synaptic currents for a representative quadripartite synapse system. (a) Action potential at the glutamatergic presynaptic nerve terminal, (b) Action potential at the GABA presynaptic nerve terminal, (c) \ac{EPSC}, and (d) \ac{IPSC}.}
    \label{fig:neuronal_spiking_activity}
\end{figure}

\begin{figure}[h!]
    \centering
    \makebox[\textwidth]{        \adjustbox{max width=\textwidth+3cm}{
            \includegraphics[width=1.2\linewidth]{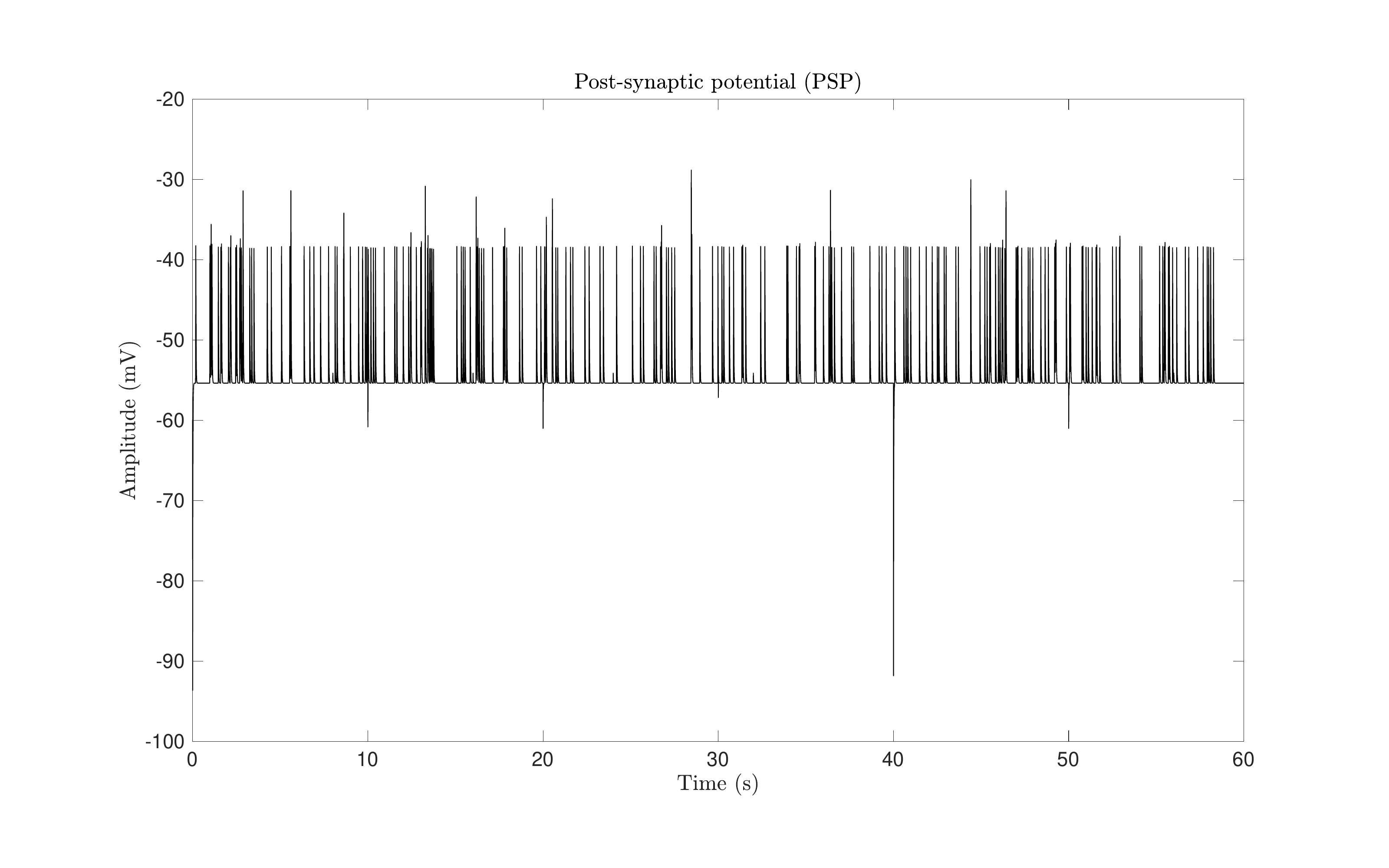}
        }
    }
    \caption{\ac{PSP} for a representative quadripartite synapse system. The PSP reflects the combined effects of excitatory and inhibitory inputs.}
    \label{fig:post_synaptic_potential}
\end{figure}

The neuronal spiking activity associated with a representative quadripartite synapse system is shown in Figures~\ref{fig:neuronal_spiking_activity} and~\ref{fig:post_synaptic_potential}. These figures provide an in-depth view of the electrical dynamics and synaptic currents that govern the interaction between excitatory and inhibitory components in the synapse.

Figure~\ref{fig:neuronal_spiking_activity} illustrates the temporal evolution of action potentials and synaptic currents. Panel (a) depicts the action potential at the \ac{Glu} presynaptic nerve terminal, characterized by rapid depolarization and repolarization cycles. This reflects the excitation dynamics triggered by \ac{Glu} neurotransmitter release, essential for synaptic activation and downstream signaling. Panel (b) shows the action potential at the \ac{GABA} presynaptic nerve terminal, where the amplitude and timing differ from those of the \ac{Glu} terminal, highlighting the distinct role of inhibitory neurotransmission in modulating the overall synaptic response.

Panel (c) presents the \ac{EPSC}, which results from the cumulative release of \ac{Glu} neurotransmitters into the synaptic cleft and their binding to post-synaptic receptors. The \ac{EPSC} curve demonstrates the temporal summation of excitatory inputs, driving the post-synaptic neuron toward depolarization. In contrast, panel (d) displays the \ac{IPSC}, generated by the release of \ac{GABA} neurotransmitters. The \ac{IPSC} curve represents the inhibitory effect on the post-synaptic neuron, counteracting excitatory signals and maintaining neuronal balance.

Figure~\ref{fig:post_synaptic_potential} focuses on the \ac{PSP}, which integrates the effects of excitatory and inhibitory inputs at the post-synaptic neuron. The PSP curve exhibits the combined depolarization and hyperpolarization phases, illustrating how synaptic inputs modulate the membrane potential and influence the likelihood of action potential generation. The dynamics of the PSP are critical for understanding the interplay between excitation and inhibition in synaptic activity and their implications for downstream neuro-glial interactions.

These results provide a comprehensive overview of neuronal spiking activity and its role in regulating synaptic responses within the quadripartite synapse system. By capturing both excitatory and inhibitory dynamics, this analysis highlights the intricate balance required for effective neurotransmission and sets the stage for further exploration of astrocytic and vascular responses in the neuro-glial-vascular coupling framework.

\subsection{Astrocytic Calcium and Gliotransmitter Signaling}
Astrocytes serve as pivotal mediators within the \ac{NGVU}, responding to neuronal activity with intricate Ca$^{2+}$ and \ac{IP$_3$} signaling cascades. These signals are not only essential for regulating gliotransmitter release but also play a critical role in vascular responses.

\begin{figure}[h!]
    \centering
    \makebox[\textwidth]{        \adjustbox{max width=\textwidth+3cm}{
            \includegraphics[width=1.4\linewidth]{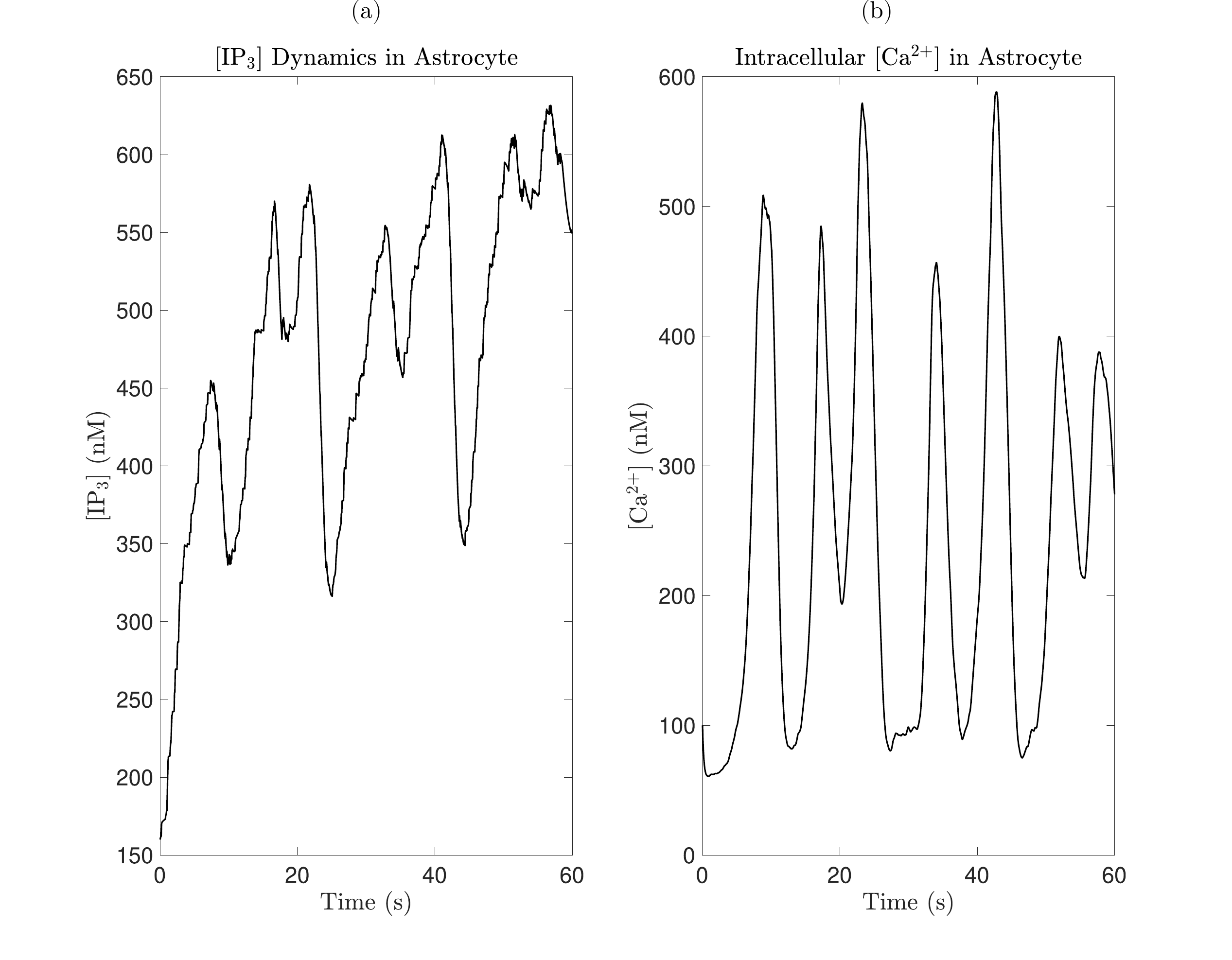}
        }
    }
    \caption{Astrocytic signaling dynamics. (a) [IP$_3$] concentration in astrocytes as a response to neuronal activation. (b) Intracellular calcium concentration, showing the biphasic nature of calcium signaling in astrocytes.}
    \label{fig:astrocyte_signaling_results}
\end{figure}

Figure~\ref{fig:astrocyte_signaling_results} illustrates the temporal evolution of astrocytic [IP$_3$] and [Ca$^{2+}$] concentrations. The left panel shows the dynamics of [IP$_3$], a key second messenger involved in intracellular calcium release. The rapid increase in [IP$_3$] upon neuronal activation triggers calcium release from intracellular stores, as reflected in the right panel. The [Ca$^{2+}$] dynamics exhibit an initial sharp rise followed by a slower decay, characteristic of astrocytic calcium signaling. This biphasic pattern reflects the interplay between [IP$_3$]-mediated release and cellular mechanisms for calcium sequestration and extrusion.

Astrocytic calcium signaling underpins gliotransmitter release, facilitating crosstalk between neurons and the vascular system. By capturing these dynamics, our model emphasizes the astrocyte's role in neurovascular coupling, where gliotransmitters regulate blood vessel tone and metabolic support for active neurons.

\subsection{Neurovascular Response}

\begin{figure}[h!]
    \centering
    \makebox[\textwidth]{        \adjustbox{max width=\textwidth+3cm}{
            \includegraphics[width=1.2\linewidth]{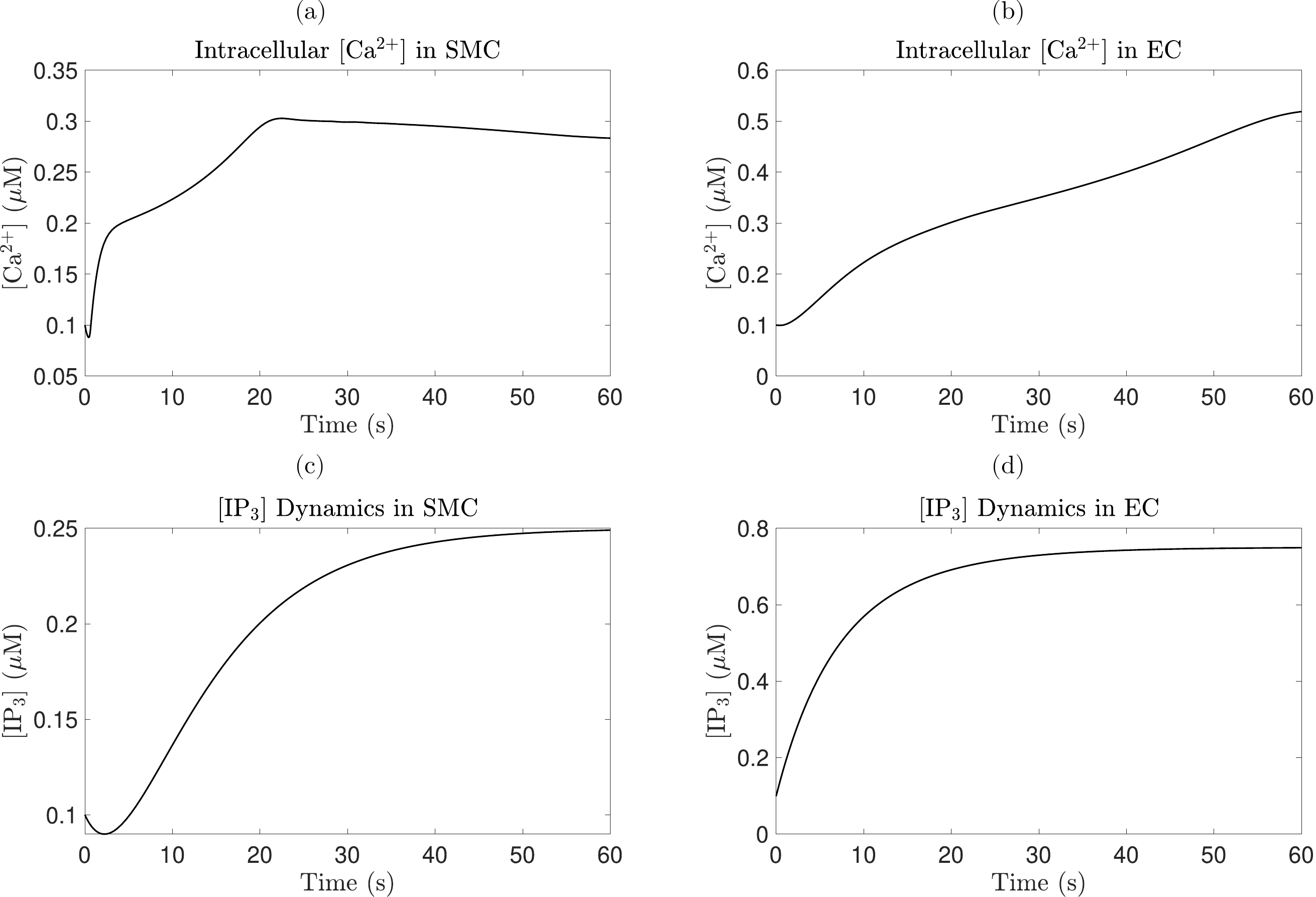}
        }
    }
    \caption{Calcium and IP$_3$ dynamics in vascular cells: (a) and (b), intracellular calcium concentration in \ac{SMCs} and endothelial cells, alongside (c) and (d), IP$_3$ signaling dynamics in these cells, highlighting the molecular interplay driving vascular responses.}
    \label{fig:vascular_signaling_results}
\end{figure}

Figure~\ref{fig:vascular_signaling_results} illustrates the temporal dynamics of calcium and inositol trisphosphate (IP$_3$) signaling within vascular \ac{SMCs} and \ac{ECs}, key components in the neurovascular coupling process. The intracellular calcium concentration in \ac{SMCs} exhibits oscillatory behavior over time, which is critical for regulating vascular tone. These calcium dynamics mediate vasoconstriction and vasodilation by modulating the contractile state of \ac{SMCs}, thereby directly influencing the radius of the blood vessel. Similarly, the intracellular calcium signals in \ac{ECs} play a complementary role by driving the release of vasoactive mediators such as \ac{NO} and prostacyclin. These mediators propagate the effects of endothelial activation to adjacent \ac{SMCs}, ensuring coordinated vascular responses.

The IP$_3$ signaling pathways, shown in the same figure, highlight the molecular mechanisms underlying calcium mobilization. In \ac{SMCs}, IP$_3$ variations reflect the activation of G-protein-coupled receptors and subsequent release of calcium from intracellular stores, such as the sarcoplasmic reticulum. In \ac{ECs}, IP$_3$ signaling serves as a critical intermediary, linking upstream signaling events to calcium mobilization and facilitating the generation of calcium waves along the endothelium. These signaling dynamics underscore the intricate interplay between \ac{SMCs} and \ac{ECs}, which is essential for maintaining vascular homeostasis and adapting to local metabolic demands.

\begin{figure}[h!]
    \centering
    \makebox[\textwidth]{        \adjustbox{max width=\textwidth+3cm}{
            \includegraphics[width=1.2\linewidth]{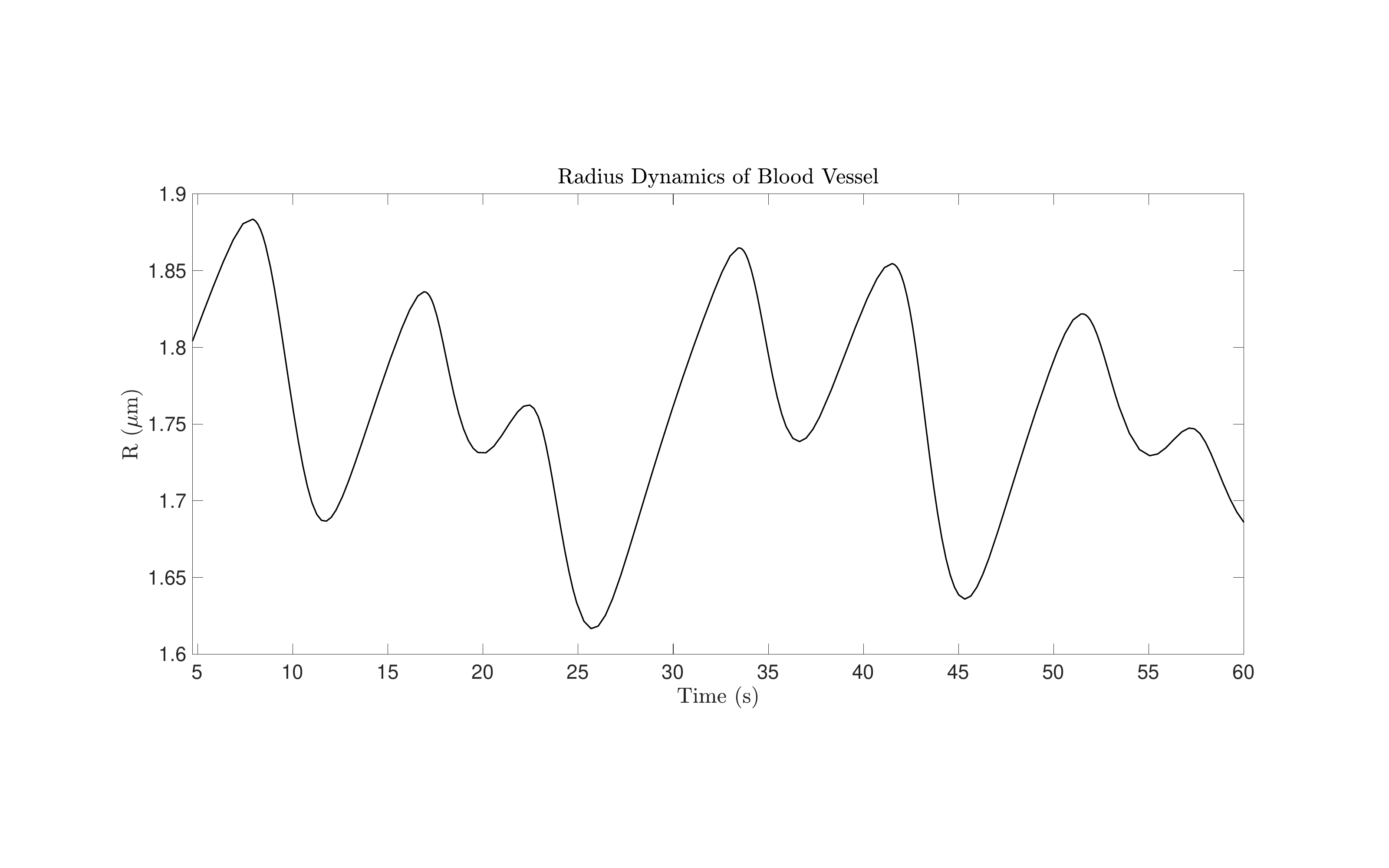}
        }
    }
    \caption{Temporal changes in blood vessel radius under neurovascular coupling dynamics. The vessel radius is allowed to settle during the first 5 seconds before dynamic fluctuations occur. These changes illustrate the interaction between intracellular signaling and vascular tone regulation, exemplifying the capillary vessel where the arterial inflow is provided.}
    \label{fig:radius_dynamics_results}
\end{figure}

Figure~\ref{fig:radius_dynamics_results} presents the dynamic changes in blood vessel radius as a result of neurovascular coupling. The variations in vessel radius reflect the synergistic effects of calcium and IP$_3$ signaling on vascular tone regulation. Vasodilation and vasoconstriction are observed in response to these intracellular signals, demonstrating the capacity of the vascular network to adapt to changes in neuronal and glial activity. This response ensures efficient oxygen and nutrient delivery to tissues, particularly in metabolically active regions. 

The results depicted in these figures validate the integrated neurovascular model by reproducing key physiological behaviors observed in experimental studies. The coupling between neuronal activation, astrocytic signaling, and vascular responses provides a comprehensive framework for understanding neurovascular coupling mechanisms and their implications in health and disease.

\subsection{Microcirculation on a Representative Unit of the \ac{DVC}}
Our computations extend the geometry of the \emph{brain99} dataset to simulate the electromechanical coupling between neuro-glial activity and vascularization within the \ac{DVC}. This model specifically captures the interaction between astrocytic and neuronal activity and the resulting dynamic changes in blood vessel radii. Blood flow dynamics and oxygen transport are analyzed within the vascular network to assess the influence of neuro-glial signals on vascular responses. The results demonstrate how capillary networks adapt to facilitate metabolic activity and neuronal function, while the hierarchical vascular structure ensures efficient blood supply and flow redistribution. This study particularly highlights the integration of macro- and microcirculation models to validate the mechanistic framework for neuro-glial-vascular coupling in the \ac{DVC}.

\begin{figure}[h!]
    \centering
        \begin{minipage}{0.45\textwidth}
        \centering
        \includegraphics[width=0.85\linewidth]{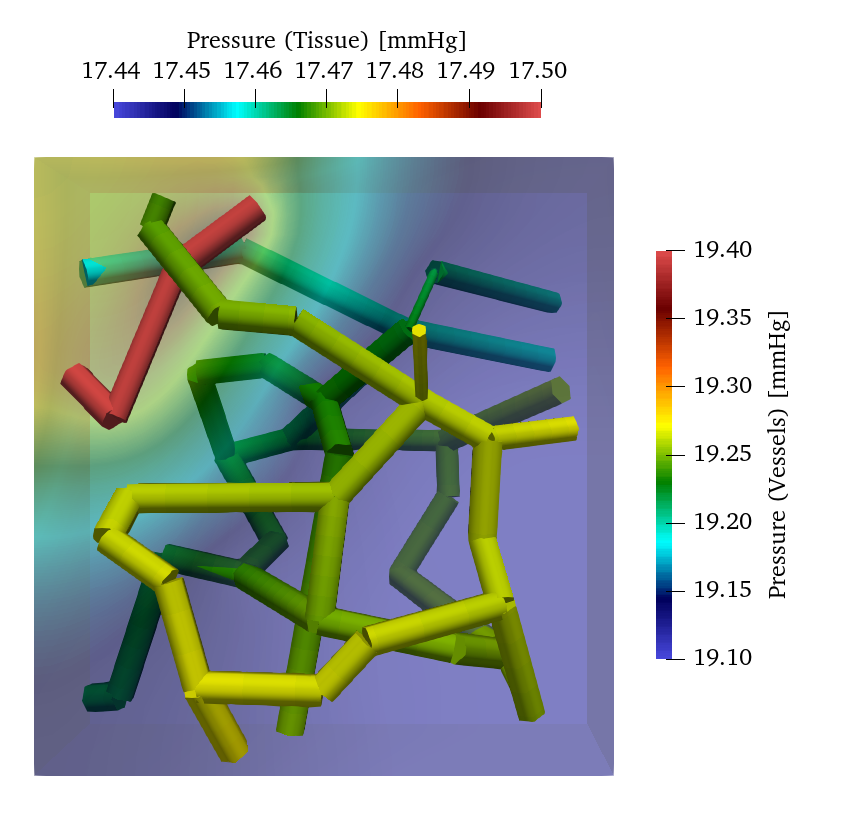}
        \\$t = 10$ s
    \end{minipage}
    \hfill
    \begin{minipage}{0.45\textwidth}
        \centering
        \includegraphics[width=0.87\linewidth]{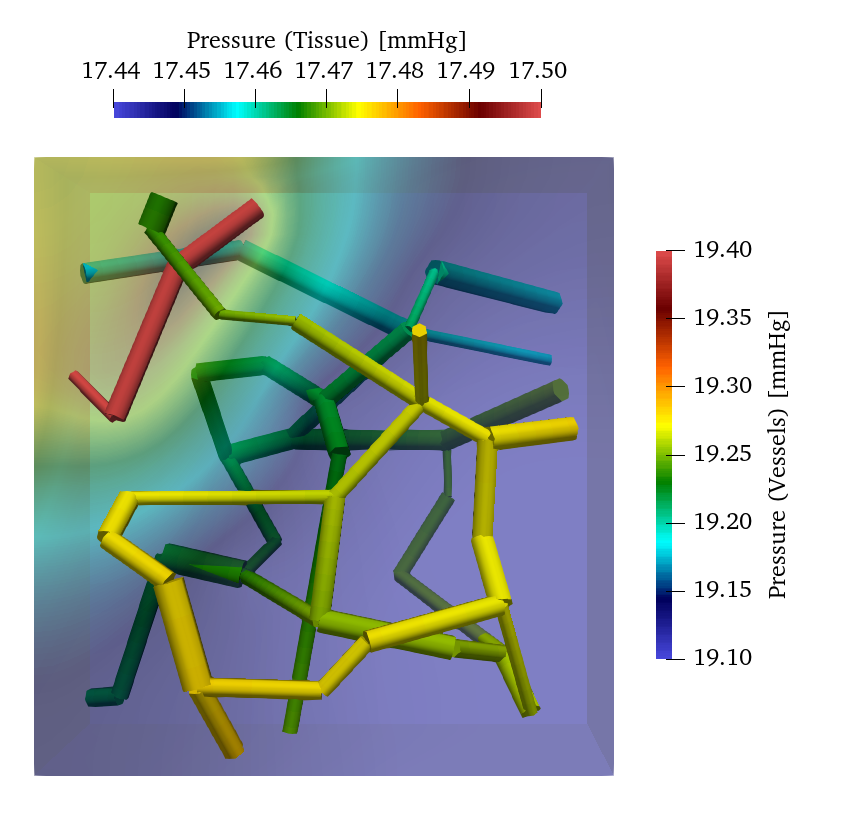}
        \\$t = 20$ s
    \end{minipage}
    \vspace{1em}
        \begin{minipage}{0.45\textwidth}
        \centering
        \includegraphics[width=0.87\linewidth]{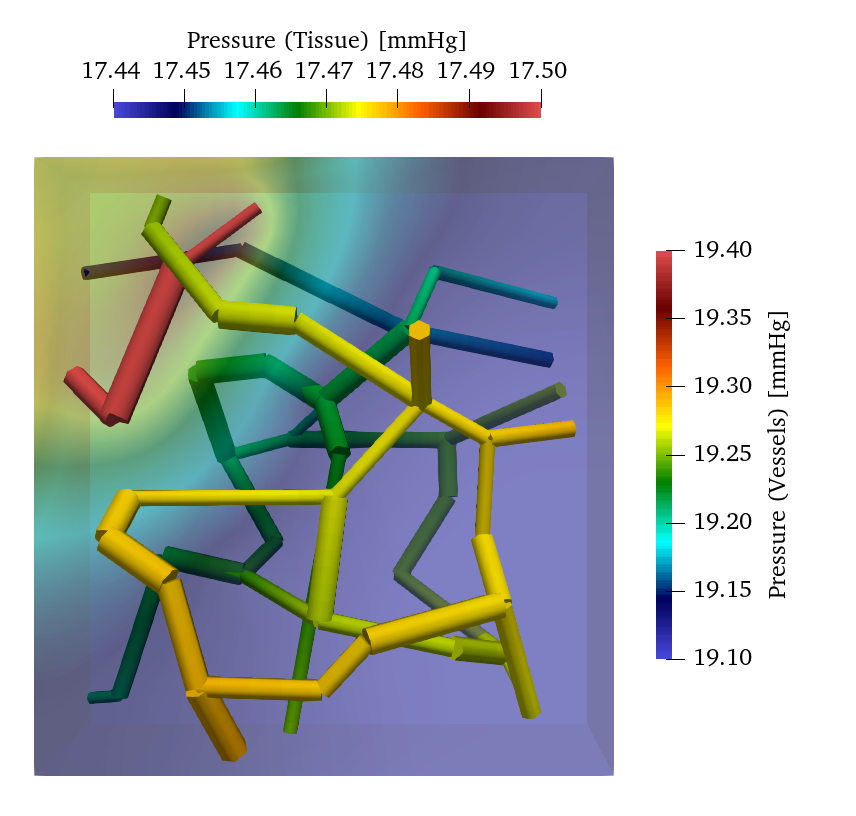}
        \\$t = 30$ s
    \end{minipage}
    \hfill
    \begin{minipage}{0.45\textwidth}
        \centering
        \includegraphics[width=0.87\linewidth]{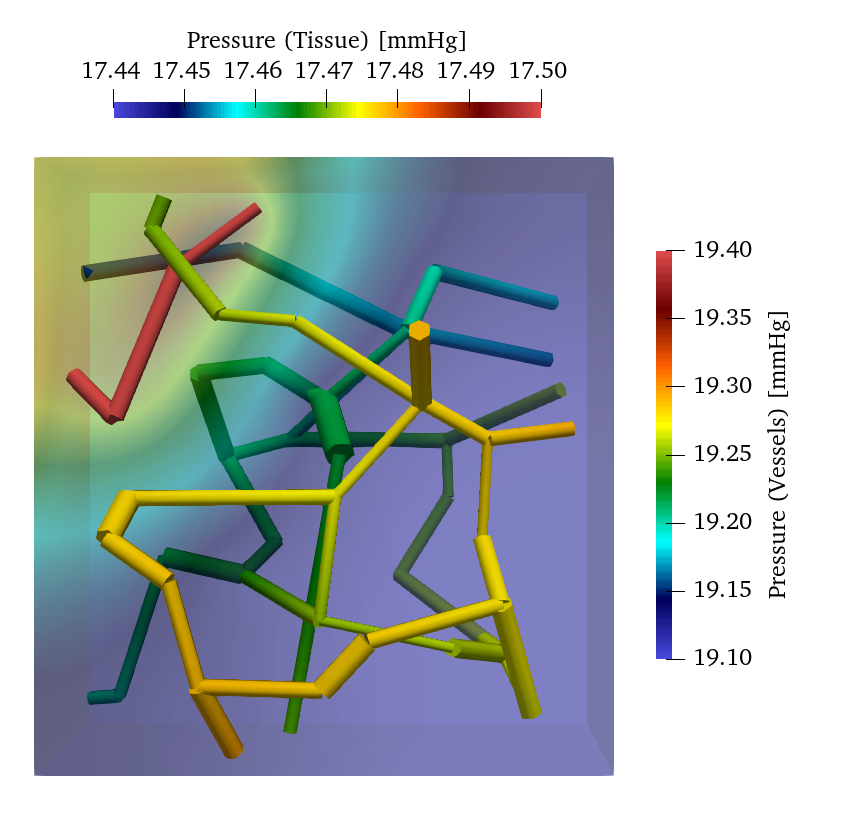}
        \\$t = 40$ s
    \end{minipage}
    \vspace{1em}
        \begin{minipage}{0.45\textwidth}
        \centering
        \includegraphics[width=0.87\linewidth]{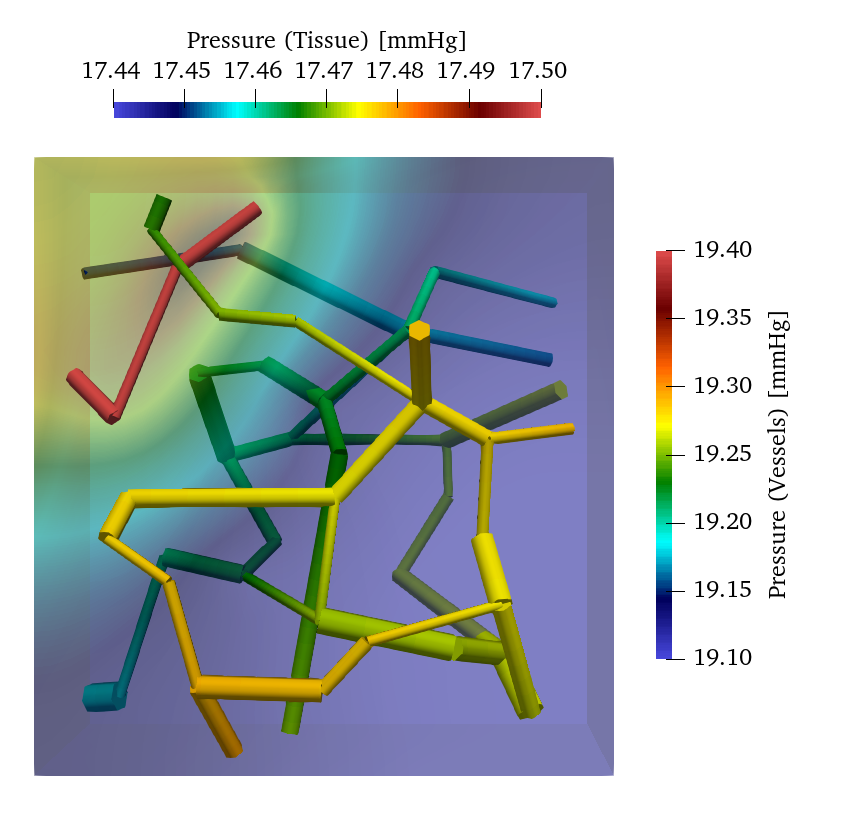}
        \\$t = 50$ s
    \end{minipage}
    \hfill
    \begin{minipage}{0.45\textwidth}
        \centering
        \includegraphics[width=0.87\linewidth]{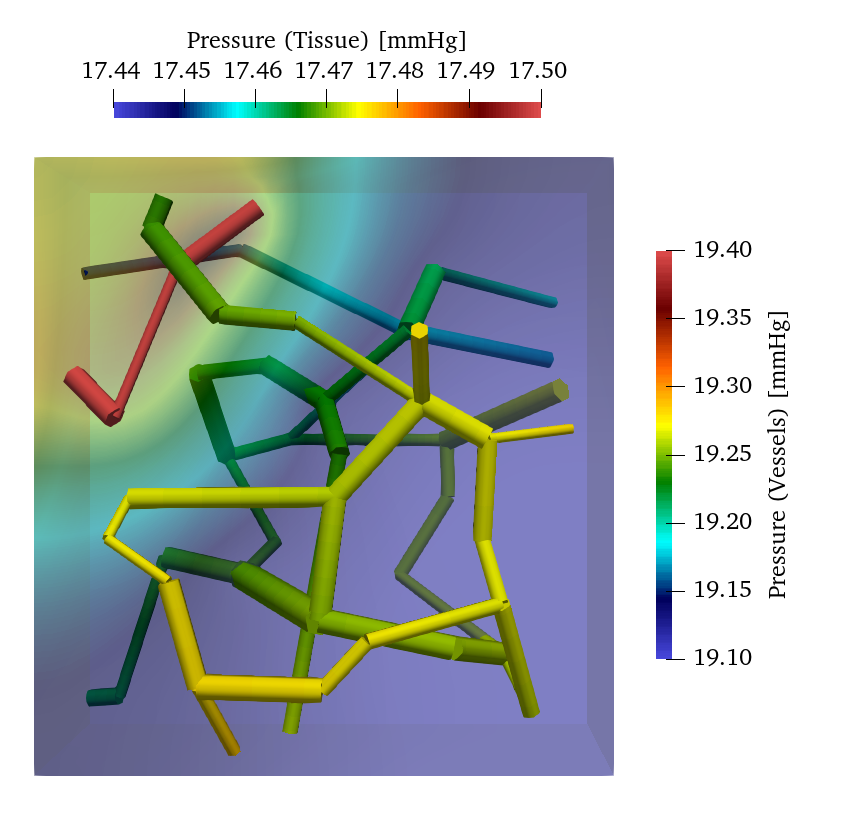}
        \\$t = 60$ s
    \end{minipage}
    \caption{Time-resolved simulation results of blood flow within the \ac{DVC} microcirculation network. A cross-sectional view is obtained by slicing the computational unit (a unit cube) along its central horizontal plane. The color map represents the local blood flow velocity magnitude, highlighting how neuro–glial–vascular coupling causes dynamic redistribution of flow via vessel dilation and constriction in both capillaries and larger vessels.}
    \label{fig:blood_flow_2x3}
\end{figure}

\begin{figure}[h!]
    \centering
        \begin{minipage}{0.45\textwidth}
        \centering
        \includegraphics[width=0.85\linewidth]{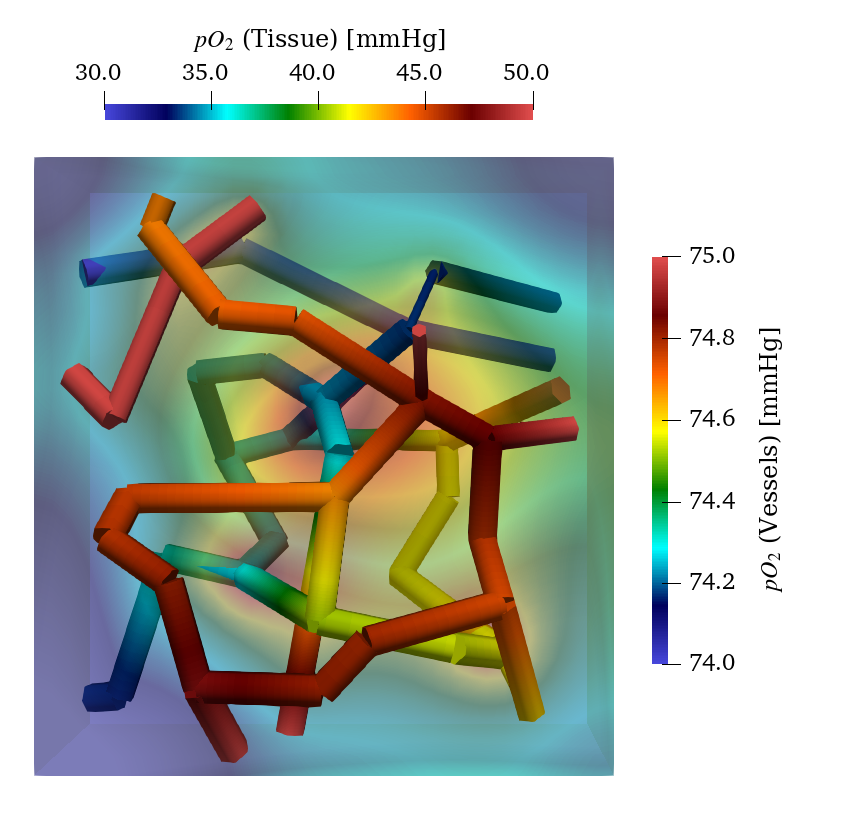}
        \\$t = 10$ s
    \end{minipage}
    \hfill
    \begin{minipage}{0.45\textwidth}
        \centering
        \includegraphics[width=0.85\linewidth]{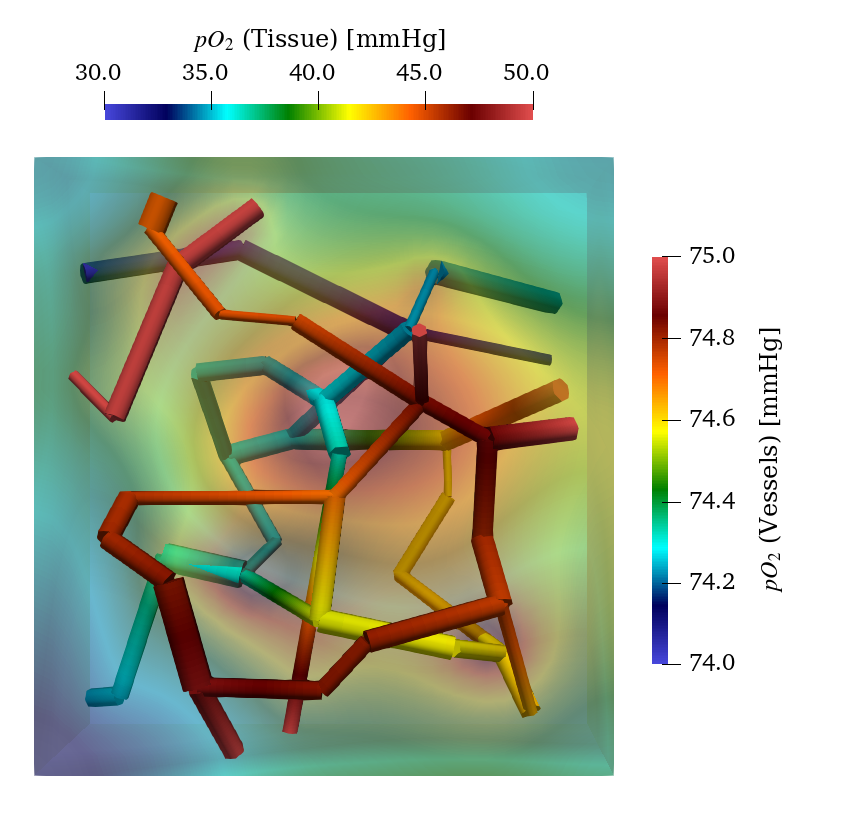}
        \\$t = 20$ s
    \end{minipage}
    \vspace{1em}
        \begin{minipage}{0.45\textwidth}
        \centering
        \includegraphics[width=0.85\linewidth]{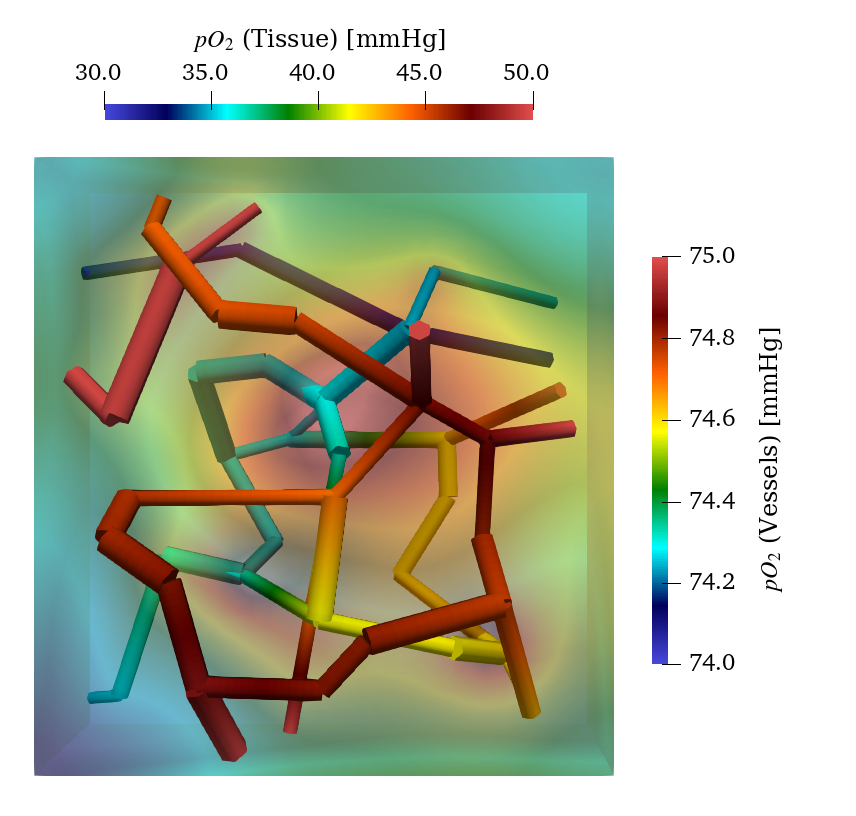}
        \\$t = 30$ s
    \end{minipage}
    \hfill
    \begin{minipage}{0.45\textwidth}
        \centering
        \includegraphics[width=0.85\linewidth]{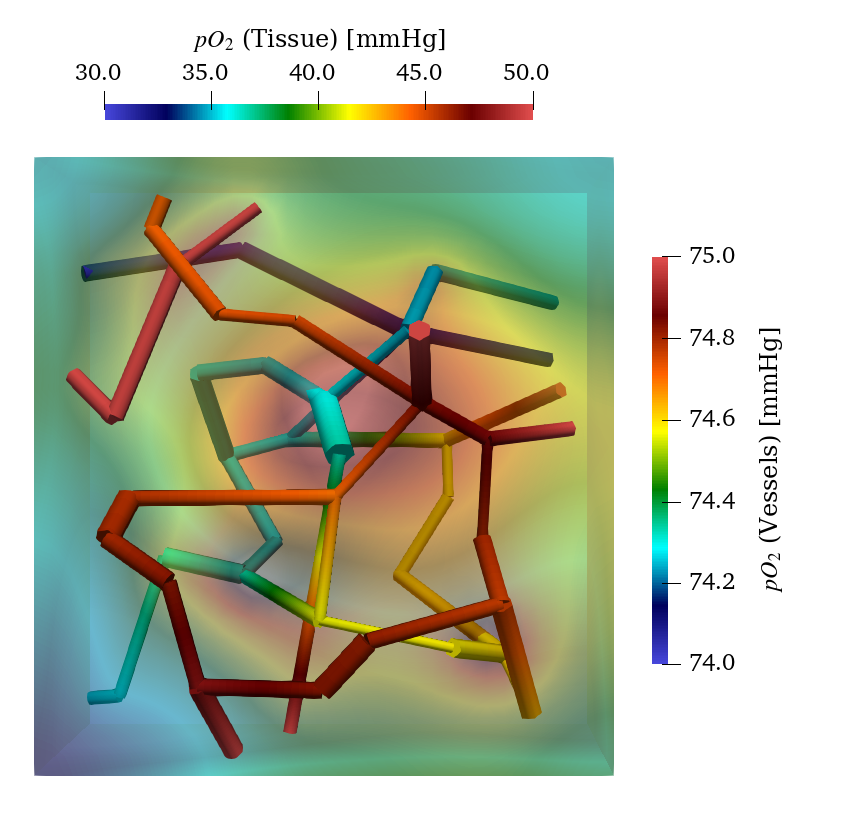}
        \\$t = 40$ s
    \end{minipage}
    \vspace{1em}
        \begin{minipage}{0.45\textwidth}
        \centering
        \includegraphics[width=0.85\linewidth]{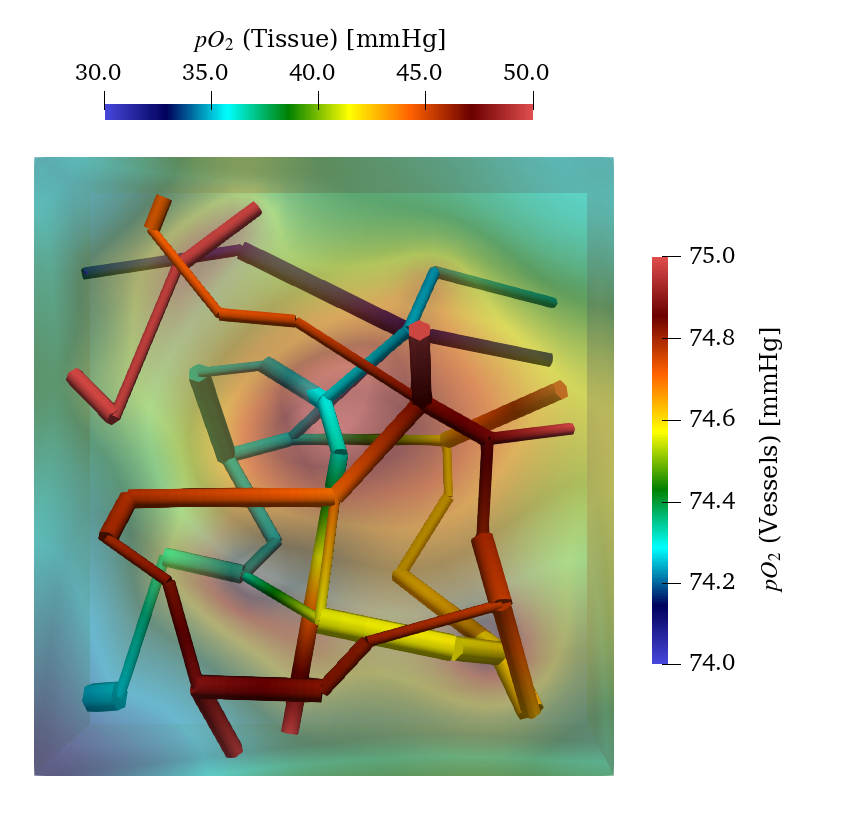}
        \\$t = 50$ s
    \end{minipage}
    \hfill
    \begin{minipage}{0.45\textwidth}
        \centering
        \includegraphics[width=0.85\linewidth]{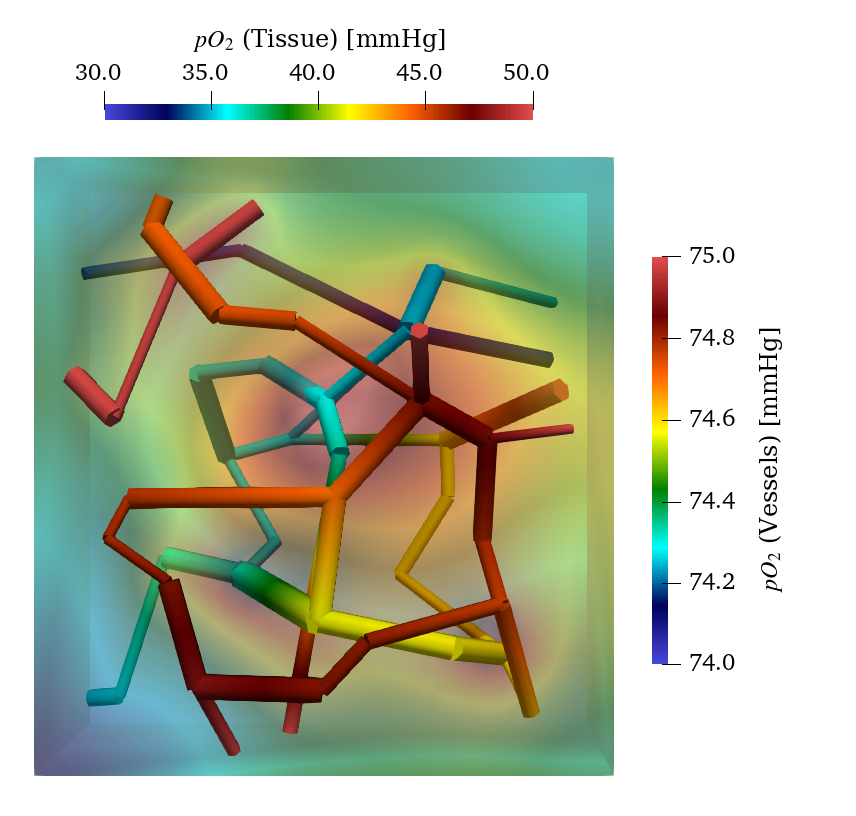}
        \\$t = 60$ s
    \end{minipage}
    \caption{Time-resolved simulation results of oxygen distribution within the \ac{DVC} microcirculation network. A cross-sectional slice through the center of the computational domain (unit cube) is rendered in 3D, with colors indicating the oxygen partial pressure ($pO_2$) in blood vessels and the surrounding tissue. This view captures the dynamics of oxygen delivery and diffusion via the 1D–3D coupling model, illustrating the impact of vessel diameter changes driven by neuro–glial interactions.}
    \label{fig:oxygen_distribution_2x3}
\end{figure}

The figures below illustrate the temporal evolution of blood flow patterns (Figure~\ref{fig:blood_flow_2x3}) and oxygen distribution (Figure~\ref{fig:oxygen_distribution_2x3}) within the \ac{DVC} microvascular network. These results reveal spatiotemporal variations in hemodynamics and oxygen delivery, computed based on the coupled neuro-glial-vascular framework. Figure~\ref{fig:blood_flow_2x3} highlights how neuro-glial activity drives dynamic changes in blood flow, with vessel dilation and constriction contributing to flow redistribution across the network. Similarly, Figure~\ref{fig:oxygen_distribution_2x3} visualizes oxygen transport and diffusion within the vascular network and surrounding tissue, emphasizing the role of capillaries in ensuring adequate oxygenation. Together, these simulations provide detailed insights into the dynamic interplay between neuronal activation, astrocytic metabolism, and vascular responses within the \ac{DVC}.
\section{Discussion}\label{sec:discussion}
The numerical results presented in this work illustrate how the integration of neuronal activity, astrocytic signaling, and vascular dynamics can yield a comprehensive understanding of neuro-glial-vascular interactions in the \ac{DVC}. By simulating the interplay between excitatory and inhibitory neurotransmitters, neuron-astrocyte calcium signaling, and blood vessel tone regulation, our model captures key physiological features of functional hyperemia, highlighting the system’s ability to meet local metabolic demands. The seamless coupling of synaptic events, astrocytic gliotransmitter release, and vascular responses underscores the intricate regulatory feedback loops that govern cerebral blood flow in health and disease.

Despite these promising findings, certain limitations must be acknowledged. The current study relies on simplified geometries and assumptions for macrocirculation, and we have primarily focused on the microcirculatory block derived from the \emph{brain99} dataset. While this approach allows for computational tractability, future models should incorporate more detailed anatomical reconstructions and account for heterogeneities at multiple scales (e.g., morphological variations in larger vessels or region-specific neuronal phenotypes). Additionally, improving the fidelity of cell-specific kinetics and including metabolic constraints—such as oxygen and glucose availability—could further enhance the model’s predictive power. These refinements would help bridge gaps between \emph{in silico} analyses and \emph{in vivo} observations, facilitating better translation to clinical and experimental contexts.

\section{Conclusion}\label{sec:conclusion}
In conclusion, we have developed a modular, multi-scale modeling framework that unifies neuronal spiking, astrocytic calcium dynamics, and vascular tone regulation within the dorsal vagal complex. Our work introduces a coupled multi-scale model that provides a seamless transition from systemic hemodynamics to local capillary dynamics. In this framework, models for flow and transport of different dimensions are coupled, resulting in a 3D-1D-0D system. By embedding a quadripartite synapse model and linking it to a macro-/microcirculatory description, we demonstrate that neuro-glial-vascular coupling emerges through highly interactive feedback processes. The proposed approach provides a valuable and novel platform for studying functional hyperemia, investigating pathological disruptions of neurovascular homeostasis, and designing future experiments. Importantly, this work represents the first integrated model of its kind for the \ac{DVC}, quantitatively capturing the interplay among neural, glial, and vascular components. Our simulation results reproduce physiologically realistic hemodynamic and synaptic behaviors in accordance with established textbook data, providing an initial validation of the model. We envisage that further refinement and extension of this model, for example by including patient-specific characteristics such as receptor heterogeneity, will offer deeper insights into cerebrovascular regulation and guide novel therapeutic strategies for various neurological conditions. Moreover, our simulation results establish a robust computational foundation for future translational investigations into both normal and pathological brain function. Finally, the generality of our methodology indicates that a similar modeling framework can be applied to other cortical areas, further broadening the scope of neurovascular research.
\section*{Acknowledgments}\label{sec:Acknowledgments}
This work was supported by the Deutsche Forschungsgemeinschaft (DFG, German Research Foundation) under project numbers 470246804 and 469698389, and through the International Graduate School of Science and Engineering (IGSSE) under project numbers WO 671/11-1 and WO 671/20-1. This research was supported by the IDIR-Project (Digital Implant Research), a cooperation financed by Kiel University, University Hospital Schleswig-Holstein and Helmholtz Zentrum Hereon.

\section*{Declarations}
The authors declare that they have no conflict of interest to disclose.

\bibliographystyle{plain}
\bibliography{refs}

\appendix

\end{document}